\documentclass[%
preprint,
amsmath,amssymb,
aps,showkeys
]{revtex4-2}
\usepackage{graphicx}
\usepackage{dcolumn}
\usepackage{bm}
\usepackage{amsmath}
\usepackage{amssymb}
\usepackage{textcomp}
\usepackage{graphicx}
\usepackage{caption}
\usepackage{subcaption}
\usepackage{array}
\usepackage{mathtools}
\usepackage{centernot}
\usepackage{slashed}
\usepackage{hyperref}
\usepackage{xcolor}
\usepackage{tikz-feynman}

\begin{document}
\title{Magnetic moment of neutrinos in a left-right symmetric model and Interplay of type-I and type-II seesaw}
\author{Vivek Banerjee}
\email{vivek\_banerjee@nitrkl.ac.in}
\author{Sasmita Mishra}
\email{mishras@nitrkl.ac.in}
\affiliation{%
Department of Physics and Astronomy, National Institute of
Technology Rourkela, Sundargarh, Odisha, India, 769008}%
\date{\today}
	
\begin{abstract}
In left-right symmetric models, the Majorana coupling 
matrix, $f$ and hence the right-handed neutrino (RHN) mass matrix 
admits eight solutions assuming the form of Dirac coupling
matrix is known. Additionally the
coupling matrix depends on the parity breaking scale, $v_R$ 
as a new physics scale.
RHNs being Majorana in nature can posses
transition magnetic moment (TMM). Neutrino magnetic moments 
are inherently related to neutrino masses as neutrino masses
imply neutrino magnetic moments.
We study, along with small neutrino
TMM, the heavy RHN transition magnetic moment contributions
to muon $g-2$, $(g-2)_\mu$ anomaly for all eight solutions of $f$. We find,
of the eight solutions, only two solutions of $f$ matrix
only contribute to the  $(g-2)_\mu$ in the experimental
predicted range. The range of $v_R$ in these cases is found to be
$3.4 \times 10^3 - 1.5 \times 10^4$  GeV. 
For a complementary check, we also study TMM induced
neutrinoless double beta decay ($0\nu\beta\beta$), for the same
set of choice of parameters. 
While certain parameter choices allow RHNs to
explain the $(g-2)_\mu$ anomaly, the same configurations lead
to an extremely long half-life for $0\nu\beta\beta$ decay, 
well beyond experimental reach. Even under extreme magnetic 
field enhancements, the half-life decreases only marginally,
reinforcing the dominance of weak interaction vertices over TMM 
contributions in $0\nu\beta\beta$ decay.
\end{abstract}
	
\keywords{Left-right symmetric model, $(g - 2)_\mu$, 
transition magnetic moment of neutrinos, neutrinoless 
double beta decay.}
	
\maketitle
\newpage   
	
\section{Introduction}
\label{sec:intro}
Originally proposed to explain the parity violation in Standard Model (SM), 
the left-right symmetric models (LRSM) \cite{Mohapatra:1974gc, Senjanovic:1975rk} are 
predictive in various ways. The gauge  group of the LRSM, 
$SU(3)_C\times SU(2)_L\times SU(2)_R \times U(1)_{B - L}$, 
is a very simple extension of the SM gauge group. To account 
for the spontaneous parity violation, the models demand the
addition of  triplet scalars and right-handed neutrinos over
and above the SM particle constituents.  The addition of new
particles in turn generate non-zero neutrino masses via type-I 
\cite{Minkowski:1977sc,Ramond:1979py,Gell-Mann:1979vob, 
Sawada:1979dis,Levy:1980ws} 
and type-II  \cite{Magg:1980ut, Lazarides:1980nt, Mohapatra:1980yp, 
Schechter:1980gr} seesaw mechanisms. It is very rewarding to observe 
that the scale of parity violation, a new scale as compared to 
that of SM, is also related to the light neutrino masses. 
Whether SM or LRSM, determining flavor structure of the model is a prime 
example of experiment driven theoretical effort. In 
SM the flavour structure resides
in the Yukawa sector of the SM Lagrangian in terms of Dirac Yukawa 
coupling of the charged fermions with the Higgs scalar (see for example 
\cite{Zupan:2019uoi}).
For neutrinos, making a similar prediction for its 
Dirac/Majorana coupling structure is not possible in SM. 
In LRSM, however,  the Dirac 
Yukawa coupling of the neutrinos can in principle be  determined 
and experimentally constrained as 
proposed by the authors of 
references \cite{Nemevsek:2012iq} and \cite{Senjanovic:2018xtu}. 
Unlike Dirac type coupling,  Majorana type coupling can not be
constrained experimentally.
In this work we sought to constrain the Majorana coupling in
LRSM using electromagnetic property of the RHNs,
considering the parity breaking scale, $v_R$ as the scale of 
new physics. We consider the latest results from the 
experiments based on anomalous magnetic moment (AMM) of muon, 
$(g-2)_\mu$ and half-life of $0\nu\beta\beta$ decay for the phenomenological study.

Although Casas-Ibarra parametrization \cite{Casas:2001sr} can determine 
the Dirac Yukawa coupling in a type -I seesaw framework,  the presence of
an arbitrary orthogonal matrix introduces some ambiguity and the knowledge of 
the origin of neutrino mass remains unknown.
In LRSM, type-I and type-II contributions to light neutrino mass are
inherently related as both contributions depend on the same
Majorana type coupling matrix.
In phenomenological studies based on LRSM, one often assumes the dominance of
either type-I or type-II seesaw contribution to light  neutrino mass, while
assuming specific values of unknown seesaw parameters. But considering both 
contributions comparable, 
the interplay of type-I and type-II seesaw can have interesting implications
for the interpretation of neutrino data. 
The seesaw formula including
both type-I and type-II contributions can be employed to reconstruct 
the Majorana coupling matrix, taking certain quantities as input 
parameters \cite{Akhmedov:2006de}. Considering both contributions 
comparable, the authors of Ref. \cite{Akhmedov:2006de}
show that the seesaw equation in LRSM has eight sets of solutions 
for the eigenvalues of the  Majorana coupling matrix, $f$.
In LRSM, the eigenvalues of the  mass matrix of the right-handed 
neutrinos, $M_R = v_R f$, hence allow eight solutions. Thus the 
study can provide some insight into the underlying theory at the
seesaw scale. 
In a similar direction the authors of Ref.\cite{Hosteins:2006ja} propose a systematic procedure for reconstructing the eight
solutions for the matrix $f$ and apply
the procedure to a particular class of supersymmetric $SO(10)$ model. 
In our previous study \cite{Banerjee:2023aro}, using the approach 
developed in Ref.\cite{Hosteins:2006ja},
we have studied the role of eight different solutions of the 
matrix $f$, by considering the present and future 
sensitivity of searches of $0\nu\beta\beta$ decay. In Ref. 
\cite{Banerjee:2023aro}, we have shown a possibility of new 
physics contributions saturating the experimental 
bound on effective mass for $v_R$ in the range of $10$ TeV for 
two particular solutions of the Majorana coupling matrix and 
simultaneously provides the insights about parity breaking scale. 
In this work, we extend the analysis by using the magnetic moment
of RHNs to study the properties of Majorana coupling, taking 
$v_R$ as the new physics scale.

In SM, apart from weak interactions,  neutrinos
could exhibit electromagnetic properties. Being neutral,
the radiative corrections can induce finite magnetic moments 
in neutrinos \cite{Bernstein:1963qh, Pal:1981rm, PhysRevD.34.3444}. 
It can interact with the virtual charged particles (weak gauge bosons, 
charged leptons, $\nu \rightleftarrows W^{\pm} l^{\pm}$) and 
result in an induced magnetic moment \cite{Babu:1991zh}.
It has been shown in references \cite{Akhmedov:1993ta,Mikheev:1986wj,
Smirnov:1991ia}, apart from undergoing resonant flavor transitions when 
propagating through matter \cite{Wolfenstein:1977ue}, neutrinos 
may transit between
different helicities due to presence of a nonuniform magnetic
field. Thus, to obtain the oscillations of the type 
$\nu_\alpha \rightarrow \bar{\nu}_\alpha$, where $\alpha
\equiv e, \mu, \tau$, it will be driven by the interplay between 
the neutrino mixing matrix and the neutrino magnetic
moment $\mu_{\alpha\beta}$. Being neutral, the magnetic moment can
be induced in second order process.
The CPT theorem ensures that the Dirac neutrinos have both diagonal and non-diagonal magnetic moments and the Majorana neutrinos can have only transition 
(in the flavour basis) magnetic moment \cite{Giunti:2014ixa}.
With addition of RHNs to SM, the neutrinos 
develop a small Dirac mass and its 
magnetic moment is $3 \times 10^{-20} \mu_B ~(m_\nu / 0.1{\rm eV})$
\cite{Fujikawa:1980yx}, where $\mu_B = e/2m_e$ is the Bohr magneton,
$e$ and $m_e$ are charge and mass of the electron respectively. 
If the neutrinos are Majorana particles, their TMMs
are even smaller, at most of order $ \mathcal{O}(10^{-23})\mu_B$
\cite{Pal:1981rm}. The explanation of various anomalous events in
observations considering neutrino magnetic moments prefer 
substantially higher values $ \mathcal{O}(10^{-11})\mu_B$ 
\cite{Babu:1989wn, 
TEXONO:2006xds,Borexino:2017fbd}, offering a challenge 
to reach this threshold in beyond SM contexts. 
Some non-standard scenarios suggest very high neutrino magnetic 
moment $\mathcal{O}( 10^{-6}\mu_B)$  
\cite{Liu:1989me, Tarazona:2015drz} but lacks naturalness due to fine 
tuning. By adding RHNs to SM, due to their singlet nature, RHNs
can not develop magnetic moment via loop diagrams. They could develop 
it through their mixing with active neutrinos. But due to the subdue
nature of mixing, it results in extremely small magnetic moment for RHNs.

Contrary to the SM neutrinos, the RHNs can have large magnetic
moments \cite{Boyarkin:2008zz,Aparici:2009fh,Aparici:2009oua}. In LRSM
due to the gauge interaction of the RHNs with the gauge bosons, RHNs can
interact via electromagnetic interaction through the magnetic 
moment operator. So models with RHNs can contribute significantly to 
spin-flavor precision, beta decay of neutron in a magnetic field and related
phenomenology \cite{Giunti:2014ixa} with contributions from the transition
magnetic moment. An experimental perspective of magnetic moment of RHNs
can be found in references \cite{Kikuchi:2008ki,Barducci:2022gdv,Chun:2024mus}. 
Like the SM neutrinos, the magnetic moment of RHNs is also dependent 
on their masses. In our study, in a class of LRSM, we explore this feature 
of RHNs to study the properties of the eight solutions of Majorana coupling matrix
through the masses of RHNs ($M_R=fv_R$).
Moreover, to contrast with experimental results we consider
the AMM of muon and half-life of $0\nu \beta \beta$ as the phenomenological study.
We express the experimental quantities of interest as a function of RHN masses
and analyse the properties of eight solutions of Majorana coupling matrix
as a function of $v_R$. 

The paper is organised as follows. Section (\ref{sec:mlrsm}) delves into 
summarizing briefly the model based on left-right symmetry,
providing required insights into its structure including field 
content, neutrino mass matrix, Majorana coupling 
matrix, and the magnetic moment of neutrinos.
In section (\ref{sec:amm}), a comprehensive theoretical framework is 
established for understanding the anomalous magnetic moment, employing 
the concept of formation of form factors extracted from loop diagrams 
with TMM vertices.
Simultaneously, the $0\nu\beta\beta$ decay is
studied in section (\ref{sec:TMM}), by studying the diagrams
including TMM induced by RHNs. 
We summarize the results for the possible dominance patterns followed by
the conclusion in section (\ref{sec:discussion}).
\section{Minimal left-right symmetric model and neutrino masses}
\label{sec:mlrsm}
In comparison to SM, LRSM deliver a parity symmetric 
environment based on the gauge group, 
$SU(3)_C\times SU(2)_L\times SU(2)_R \times U(1)_{B - L}$,
where leptons and quarks are expressed as left
and right chiral doublets. 
Because of the extended gauge group, it introduces several 
new parameters corresponding to the newly added fields. 
The fermion contents of this model with their gauge charges are
given by, 
\begin{equation}
Q_L \left(3,2,1,\frac{1}{3}\right) = \begin{bmatrix} u \\ 
d  \end{bmatrix}_L , \begin{bmatrix} c \\ s  
\end{bmatrix}_L ,\begin{bmatrix} t \\ b  \end{bmatrix}_L, 
\quad Q_R \left(3,1,2,\frac{1}{3}\right) = \begin{bmatrix} u \\ d 
\end{bmatrix}_R, \begin{bmatrix} c \\ s  \end{bmatrix}_R, 
\begin{bmatrix} t \\ b  \end{bmatrix}_R,
\end{equation}
\begin{equation}
L_L \left(1, 2, 1, -1 \right) = \begin{bmatrix} \nu_e \\ 
e  \end{bmatrix}_L , \begin{bmatrix} \nu_\mu \\ \mu  
\end{bmatrix}_L ,\begin{bmatrix} \nu_\tau \\ \tau  \end{bmatrix}_L, 
\quad L_R \left( 1, 1, 2, -1 \right) = \begin{bmatrix} N_e \\ e 
\end{bmatrix}_R, \begin{bmatrix} N_\mu \\ \mu  \end{bmatrix}_R, 
\begin{bmatrix} N_\tau \\ \tau  \end{bmatrix}_R.
\end{equation}
Here $L$ and $R$ in the subscripts stand for the left-handed 
and right-handed chirality respectively. With two scalar 
triplets and a bidoublet, LRSM 
is termed as minimal left-right symmetric model (MLRSM). 
The Higgs content of this model can be represented as \cite{Deshpande:1990ip},
\begin{equation}
\phi \left(1,2, 2^*, 0 \right) = 
\begin{bmatrix} \phi_1^0 & 
\phi_1^+ \\ \phi_2^- & \phi_2^0  \end{bmatrix},\quad
\Delta_L \left( 1,3,1,2 \right) =\begin{bmatrix} \frac{\delta_L^+}{\sqrt{2}} &
\delta_L^{++} \\ \delta_L^0 & \frac{\delta_L^+}{\sqrt{2}}
\end{bmatrix},\quad
\Delta_R(1,1,3,2) =\begin{bmatrix} \frac{\delta_R^+}{\sqrt{2}} 
& \delta_R^{++} \\ \delta_R^0 & \frac{\delta_R^+}{\sqrt{2}} 
\end{bmatrix}.
\end{equation}  
The gauge symmetry of the MLRSM breaks into the SM gauge
group with the scalar fields developing a desired vacuum alignment to
have an electric charge conserving minima of the scalar 
potential \cite{Senjanovic:1978ev,Mohapatra:1979ia,Mohapatra:1980yp}, 
given by,
\begin{equation}
\langle \phi \rangle =\begin{bmatrix} \frac{k_1}{\sqrt{2}} & 0 \\ 0 & 
\frac{k_2}{\sqrt{2}}  \end{bmatrix}, \quad \langle \Delta_{L,R} \rangle=
\begin{bmatrix} 0 & 0 \\ \frac{v_{L,R}}{\sqrt{2}} & 0 
\end{bmatrix}.
\end{equation}
Taking only the effects of $v_R$ into account, it can be shown that 
$v_R$ breaks $SU(2)_R\times U(1)_{B-L}$ to $U(1)_Y$. The complete 
symmetry breaking pattern can be shown as,
\begin{equation}
SU(2)_L\times SU(2)_R\times U(1)_{B-L} \xrightarrow{v_{R}} SU(2)_L\times U(1)_Y
\xrightarrow{k_1, k_2} U(1)_{\rm em}.
\end{equation}	
The SM Higgs vacuum expectation value (vev), $v$ is related to the 
vev of the bidoublet with the relation, $v^2 = k_1^2 + k_2^2$. 
 Here the hierarchy of vevs follow the relation,
\begin{equation}
|v_L|^2 \ll |k_1^2| + |k_2^2| \ll |v_R|^2.
\end{equation}
Now, moving to the leptonic part of MLRSM, the Lagrangian containing 
the Yukawa interaction terms for leptons is,
\begin{equation}
\mathcal{L}_{\text{Yukawa}}^{\text{lepton}}=h^Y_{ij}\bar{L_L^i}
\phi L_R^j + g^Y_{ij}\bar{L_L^i}\tilde{\phi}L_R^j + 
i(f_L)_{ij}L_L^{iT}C\tau_2\Delta_L L_L^j + i(f_R)_{ij}
L_R^{iT}C\tau_2\Delta_R L_R^j + {\rm h.c}.
\label{eq:yl}
\end{equation}
Here, $h^Y_{ij}$ and $g^Y_{ij}$ are the Dirac-type Yukawa coupling 
matrices, and $f_L$, $f_R$ are the Majorana type Yukawa coupling 
matrices. The generation indices are represented as, $i, j= 1,2,3$.
The discrete left-right symmetry in the model can be implemented
either by the parity operator, $P$ or the charge conjugation 
operator, $C$. Here we consider $P$ as the operator under which 
the fields transform as,
\begin{equation}
\psi_L\longleftrightarrow \psi_R,\quad \Delta_L
\longleftrightarrow\Delta_R,
\quad \phi\longleftrightarrow \phi^\dagger.  
\end{equation}
Requiring the Lagrangian is symmetric under parity as the
left-right symmetry, the Yukawa couplings are constrained as, 
$h^Y_{ij} = h^Y_{ji}$ and $g^Y_{ij} = g^Y_{ji}$, 
and $f_L$ = $f_R$. Henceforth, we shall use $f_L =f_R =f.$

Because of the left-right symmetry, the 
charged gauge bosons have left and right counterparts too. The
corresponding Lagrangian for charged current interaction of leptons  can be written as,
\begin{equation}
\mathcal{L}_{CC}^l = \frac{g_L}{\sqrt{2}}\bar{L}_{L_i}
\gamma_{\mu}\nu_{L_j} W_L^{\mu} + \frac{g_R}{\sqrt{2}}\bar{L}_{R_i}
\gamma_{\mu}N_{R_j} W_R^{\mu} + {\rm h.c},
\end{equation} 
which shows the gauge interactions involving leptons and 
charged gauge bosons. Here, $g_L(W_L)$ and $g_R(W_R)$ are the gauge 
coupling constants (gauge bosons) of $SU(2)_L$ and $SU(2)_R$ gauge sectors 
respectively. The process of left-right
and electroweak symmetry breaking, give masses to the  weak
gauge bosons of the theory, and as a result mixing between $W_L$
and $W_R$ can take place. The two mass eigenstates 
of them are defined as $W_1$ and $W_2$, which are related to the
gauge eigenstates as,
\begin{equation}
\begin{pmatrix}
W_L \\ 
W_R 
\end{pmatrix} =
\begin{pmatrix}
\cos\zeta  & e^{i\phi}\sin\zeta \\
-e^{-i\phi}\sin\zeta & \cos\zeta
\end{pmatrix} \begin{pmatrix}
W_1 \\ W_2
\end{pmatrix},
\label{eq:mixing}
\end{equation} 
where the mixing angle $\zeta$ is represented as, $\tan~ 
(2\zeta) = \frac{2k_1 k_2}{v_R^2 - v_L^2}$ and $\phi$ is
a CP violating phase.
The mass matrix in gauge basis can be expressed as,
\begin{equation}
\begin{pmatrix}
\frac{1}{2}g_L^2(k_1^2 + k_2^2 + 2v_L^2) & g_Lg_R k_1k_2 \\
g_Lg_R k_1k_2 & \frac{1}{2}g_R^2(k_1^2 + k_2^2 + 2v_R^2)
\end{pmatrix}.
\label{eq:massofmass_st}
\end{equation}
Because of the left-right gauge symmetry, by considering 
$g_L = g_R = g$, the physical masses
of $W_1$ and $W_2$ states can be  calculated. They are given by,
\begin{equation}
\begin{split}
m_{W_1}\simeq \sqrt{\frac{1}{2}g^2(k_1^2 + k_2^2)}, 
\hspace{15pt} m_{W_2}\simeq \sqrt{\frac{1}{2}g^2(k_1^2 + 
k_2^2 + 2v_R^2)}.
\end{split}
\end{equation} 
Here, the mass of $m_{W_1}$ is approximately equal to the SM
charged weak gauge boson $W^{\pm}$ mass, whereas the $m_{W_2}$ mass
is proportional to $v_R$, the new physics scale. In the low energy 
limit, the effective mass relation between mass states and gauge
states of the charged weak gauge bosons are, 
\begin{equation}
M_{W_L}\simeq m_{W_1}, \hspace{15pt} M_{W_R}\simeq m_{W_2}.
\end{equation}
Now coming to the lepton sector, the mass matrix of the charged
lepton becomes,
\begin{equation}
m(l^-)=\frac{1}{\sqrt{2}}(h^Yk_2 + g^Yk_1).
\end{equation}
Unlike SM, in MLRSM, Dirac and
Majorana  mass terms can be generated for neutrinos. 
As for Majorana particles, particle and anti-particle fields are the
same, it is expressed as, $\psi^c = \psi; \psi^c = C\bar{\psi}^T$. So, 
taking both Dirac and Majorana fields into account, a new basis for 
the neutrino mass matrix can be formed as, $(\nu, N)^T$. Where $\nu$
and N are represented as,
\begin{equation}
\nu=\frac{\nu_L + \nu_L^c}{\sqrt{2}}, \quad N=
\frac{\nu_R + \nu_R^c}{\sqrt{2}}.
\end{equation}

After the series of symmetry breaking, the complete neutrino mass matrix 
can be written as,
\begin{equation}
		M_{\nu} = \begin{pmatrix}fv_L & y_D v \\ y_D^T v & f v_R 
		\end{pmatrix}_{6\times6} = \begin{pmatrix}
			m_L & m_D \\ m_D^T & M_R 
		\end{pmatrix}_{6\times6}; ~~
		y_D =\frac{1}{\sqrt{2}}\left(\frac{h^Yk_1 +g^Yk_2}{v}\right)
		\label{eq:mass_matrix},
	\end{equation}
in the basis of $(\nu,N)^{\rm T}$. The neutrino mass matrix is 
diagonalized using the unitary matrix, $U_{6\times6}$. The 
unitary matrix $U$ can be decomposed as $U=VU_\nu$, which has 
the form,
\begin{equation} 
\begin{split}
U =VU_\nu 
&=\begin{pmatrix} 1 -\frac{1}{2}RR^{\dagger}  & R \\ 
-R &  1 -\frac{1}{2}R^{\dagger} R \end{pmatrix} 
\begin{pmatrix} U_L & 0 \\ 0 & U_R \end{pmatrix}, \\
&=\begin{pmatrix} U_L -\frac{1}{2}RR^{\dagger} U_L & RU_R \\
-RU_L & U_R -\frac{1}{2}R^{\dagger} RU_R \end{pmatrix}
=\begin{pmatrix} U_L' & T \\ S & U_R'  \end{pmatrix}.
\end{split}
\label{eq:6c6}
\end{equation}
The matrix R used in the previous equation is defined as, 
$R=m_D^{\dagger}M_R^{-1}$. The $U_\nu$ is a block diagonal
$6\times6$  matrix, $U_{\nu} = \text{Diag}(U_L, U_R)$. 
The $U_R$ matrix is the diagonalizing matrix of the RHN mass 
matrix, which can be obtained numerically in our work,
from the diagonalizing matrix of $f$ matrix.
In the present work, we have taken $U_L$ as the
Pontecorvo-Maki-Nakagawa-Sakata (PMNS) matrix, $U_{\text{PMNS}}$ 
times a diagonal matrix involving Majorana phases, 
${\rm Diag}(1, e^{i\alpha}, e^{i\beta})$. The structure 
of $U_{\rm PMNS}$ matrix can be formed using three rotation 
matrices with corresponding rotating angles $\theta_{ij}; i,j =1,2,3$,
and a Dirac CP violating phase $\delta_{CP}$. The PMNS matrix
for light neutrino sector is,
	\begin{equation}
	\centering
	U_{\rm PMNS} = 
	\begin{pmatrix} 
	c_{12}c_{13} & s_{12}c_{13} & s_{13} e^{-i\delta_{CP}}  \\
	-s_{12}c_{23} - c_{12}s_{13}s_{23}e^{i\delta_{CP}} & c_{12}c_{23} - s_{12}s_{13}s_{23}e^{i\delta_{CP}} & c_{13}s_{23} \\ s_{12}s_{23} - c_{12}s_{13}c_{23}e^{i\delta_{CP}} & -c_{12}s_{23} - s_{12}s_{13}c_{23}e^{i\delta_{CP}} & c_{13}c_{23}
	\end{pmatrix},                  
	\end{equation}
where $c_{ij} = \cos (\theta_{ij})$, $s_{ij} = \sin (\theta_{ij})$.
After diagonalizing, the neutrino mass matrix reduces to the block diagonal form,
\begin{equation}
U^T M_{\nu} U = U^T \begin{pmatrix} m_L & m_D \\ m_D^T & M_R
\end{pmatrix} U = \begin{pmatrix}
m_{\text{Diag}} & 0 \\ 0 & M_{\text{Diag}}
\end{pmatrix},
\end{equation}
where $m_{\rm Diag}$ and $M_{\text{Diag}}$ are the diagonal matrices
of light neutrino and RHN masses respectively. The latter are related to the mass matrices in flavor basis via the relations,
\begin{equation}
 U_L^T m_\nu U_L = m_{\rm Diag}, ~~ U_R^T M_R U_R = M_{\text{Diag}},  
\end{equation}
The expressions for mass matrices for light and RHNs, in flavor basis, are given by,
\begin{equation}
m_\nu \simeq f v_L - \frac{v^2}{v_R}y_D^T f^{-1}y_D
, \quad M_R =f v_R.
\label{eq:nmlr}
\end{equation}
Here the mass for light neutrino includes two terms, which are similar
to the mass generated in seesaw type-I 
and type-II mechanisms and we will use this feature of MLRSM while 
studying the dominance patterns of Majorana type Yukawa couplings.
\subsection{Introducing the interplay between Type-I and Type-II 
seesaw dominance}
\label{ssec:dominance}
In Eq.(\ref{eq:nmlr}), the light neutrino mass $m_{\nu}$ is
sum of contributions from type-I and type-II seesaw limits. Just by looking at the light
neutrino mass expression, one can have the idea of the leading term 
(defined as the dominant term) depending on the conditions as,
\begin{equation}
f v_L \ll \frac{v^2}{v_R}y_D^T f^{-1}y_D \longrightarrow 
m_\nu \simeq  - \frac{v^2}{v_R}y_D^T 
f^{-1}y_D \quad({\text{Type-I dominance}}),
\label{eq:typeI}
\end{equation}
\begin{equation}
f v_L \gg \frac{v^2}{v_R}y_D^T f^{-1}y_D \longrightarrow m_\nu \simeq fv_L \quad 
({\text{Type-II dominance}}).
\label{eq:typeII}
\end{equation} 

In case of type-I dominance (Eq.(\ref{eq:typeI})) in light
neutrino mass the $f$ matrix can be reconstructed as a function
of $v_R$ assuming Dirac coupling is known. Similarly for 
type-II dominance (Eq.(\ref{eq:typeII})) the $f$ matrix can be
reconstructed using light neutrino mass matrix. 

Considering Eq.(\ref{eq:nmlr}), without assuming any seesaw 
dominance, in $m_\nu$, it becomes a quadratic function of $f$.
For one generation case, solving Eq.(\ref{eq:nmlr}) for $f$,
the solutions are,
\begin{equation}
\begin{split}
f_{\pm} &= \frac{m_{\nu}}{2v_L} \left[1 \pm \left(1 +
\frac{4v^2 y_{D}^2}{m_{\nu}^2}\frac{v_L}{v_R}\right)^{1/2}\right], 
\\ & = \frac{m_{\nu}}{2v_L} \left[1 \pm \left(1 + d\right)^{1/2}
\right].
\end{split}
\end{equation}
where $d = \left(\frac{4v^2 y_{D}^2}{m_{\nu}^2}\frac{v_L}{v_R}\right)$, 
is defined as the dominance factor.
For example, for the condition $|d| \ll 1$, the two solutions can be represented 
approximately as,
\begin{equation}
f_{+} \approx \frac{m_\nu}{v_L} +\frac{v^2 y_D^2}{v_R m_\nu}, 
\quad f_{-} \approx - \frac{v^2 y_D^2}{v_R m_\nu}.
\label{eq:domn}
\end{equation}
So, it is easy to identify $f_+$ and $f_-$ as the type-II and 
type-I dominant solutions respectively. Also in Eq. (\ref{eq:domn}),
for $v_L \ll m_\nu$ the solutions $f_{+}$ are not perturbative and
should be discarded and the only viable solution is 
$f_{-}$.
	
\subsubsection{For three generation of neutrinos}
\label{sssec:3gen}
To get the solution of $f$ for three generations,
the task becomes difficult  as compared to one generation.
With three generations of leptons, the mass expression 
for light neutrinos given in Eq.(\ref{eq:nmlr}),
is a relation among three $3\times 3$ dimensional matrices:
$m_\nu, y_D$ and $f$. The technique to obtain the 
solutions of $f$ are shown in references \cite{Akhmedov:2006de}
and \cite{Hosteins:2006ja}. Due to the three generations of neutrinos, 
eight different solutions ($2^n, n=\text{number of generation}$)
are possible with all possible dominance (type-I and/or II) patterns. 
In the  present work, the process used in \cite{Hosteins:2006ja} has
been adopted to obtain the solutions of $f$.
	
Assuming the Dirac Yukawa matrices to be symmetric and invertible,
Eq.(\ref{eq:nmlr}) can be expressed in terms of new parameters as
\begin{equation}
\text{T}=\lambda \text{Z} - \rho \text{Z}^{-1},
\label{eq:trans_eqn}
\end{equation}
where $\lambda=v_L$ and $\rho=\frac{v^2}{v_R}$.
The forms of T and Z can be written as,
\begin{equation}
\text{T} = y_D^{-1/2}m_\nu (y_D^{-1/2})^T, \quad \text{Z} = 
y_D^{-1/2}f (y_D^{-1/2})^T.
\end{equation}
Now, the roots $t_1,t_2,t_3$ of the matrix $T$ can be determined 
from the equation,
\begin{equation}
\text{Det}(\text{T}- t \textbf{I})=0.
\end{equation}
	
In the process one can get the diagonalizing 
matrix of $T$, such that $\text{T} = \text{P}_\text{T} 
\text{Diag}(t_1,t_2,t_3) \text{P}_\text{T}^T,$
and $\text{P}_\text{T}\text{P}_\text{T}^T =1$.
Using this, the Z matrix can also be diagonalized as,  $$\text{Z} = 
\text{P}_\text{T} \text{Diag}(z_1,z_2,z_3) \text{P}_\text{T}^T.$$
	
Now using $z_i$ ($i=1,2,3$), as the solutions of Z, 
Eq.(\ref{eq:trans_eqn}) can be written in the diagonal form as, 
\begin{equation}
t_i=\lambda z_i - \rho z_i^{-1}.
\label{sol}   
\end{equation}
The $f$ matrix can be expressed as,
\begin{equation} \label{eql}
\begin{split}
f & =y_D^{1/2}\text{Z}(y_D^{1/2})^T, \\
& =y_D^{1/2}\text{P}_\text{T} \text{Diag}(z_1,z_2,z_3)
\text{P}_\text{T}^T(y_D^{1/2})^T.
\end{split}
\end{equation}
Solving Eq.(\ref{sol}), the general solution becomes,
\begin{equation}
z_i^{\pm} = \frac{t_i \pm \text{Sign}(\text{Re}(t_i)
\sqrt{t_i^2 + 4\lambda\rho})}{2\lambda}.
\end{equation}
Now, in the limit $4\lambda\rho \ll t_i^2$ the solutions are,
\begin{align}
z_i^+ \simeq \frac{t_i}{\lambda},  \quad &   z_i^- \simeq 
\frac{-\rho}{t_i}.
\label{eq:1st}
\end{align} 
	
We can get the eight solutions for $f$ in the form, for example, 
$(+,+,+)$ by using $(z_1^+,z_2^+,z_3^+)$ and  $(-,+,-)$ by using
$(z_1^-,z_2^+,z_3^-)$ and 
so on. The $ z_i^+$ and $z_i^-$ signify the type-II  and type-I
dominance respectively. And using, $M_R= f v_R$ in Eq.(\ref{eq:nmlr}),
the heavy neutrino mass matrices 
can be constructed and their eigenvalues will represent the 
physical masses of the heavy RHNs, $M_1, M_2$, and $M_3$.
\subsection{A case study}
\label{ssec:case-study}
In order to reconstruct the Majorana coupling matrix $f$,
it requires the inputs as $y_D$ and $m_\nu$ at the seesaw scale,
as can be seen from Eq.(\ref{eq:nmlr}). 
While $m_\nu$ can be reconstructed from low energy experimental data,
the Dirac Yukawa coupling matrix is generally unknown.
One possible method is the top-down parametrization,
where the Dirac type Yukawa coupling matrix is defined as, 
$y_D = \frac{1}{v} V^{\dagger}_L\tilde{m}V_R$. Here, $V_L$
and $V_R$ are unitary matrices with unknown parameters and 
$\tilde{m}$ has arbitrary diagonal entries \cite{Barry:2013xxa}. 
But due to the nature of vast parameter space, which leads to 
inevitable fine tuning, this parametrization is not a 
preferable choice of $y_D$.

Here we adopt a top-down approach that constructs a viable
left-right symmetric theory from a small and controllable 
set of inputs at high scale. The requirement that $f_L = f_R$
and Yukawa coupling matrices are symmetric, arises naturally
not only in broad class of left-right symmetric models but
also in $SO(10)$ grand unified theory (GUT). The $SO(10)$ GUTs 
have multiple possible breaking patterns to the SM gauge 
group with intermediate left-right symmetric model as one 
of the breaking chains. In Ref. \cite{Banerjee:2023aro}, we have given a
brief summary of motivation behind the choice for $y_D$ as the up-type quark 
Yukawa matrices at GUT scale. 
To avoid the repetition of content, we refer the reader to 
section (5) and Appendix (A) of Ref. \cite{Banerjee:2023aro} or 
Ref.\cite{Hosteins:2006ja} for a detailed discussion.

The form for $y_D$, at GUT scale, can be constructed as,
\begin{equation}
y_D = U_q^T \hat{y} U_q; \quad U_q = P_u V_{CKM} P_d; \quad \hat{y}
= \text{Diag}(y_u,y_c,y_t).
\label{eq:gut-scale}
\end{equation}
Here $P_u$ and $P_d$ are the diagonal matrices which include 
the phases, ($\phi_u, \phi_c, \phi_t$) and ($\phi_d,\phi_s, 
\phi_b$) of the up- and down-type quark fields respectively.
The effect of renormalization group equation (RGE) run on $y_D$ 
between the GUT scale and seesaw scale are smaller than the 
uncertainty on the quark mass parameters at $M_{\rm GUT}$. 
So, the value of $y_D$ at seesaw scale can be taken to be 
same as the GUT scale. The Dirac Yukawa coupling values for
up, charm and top quarks ($y_u, y_c, y_t$) are taken to be 
$4.2 \times 10^{-6}, 1.75 \times 10^{-3}$ and $0.7$ 
respectively at GUT scale \cite{Hosteins:2006ja}. The 
values for Yukawa couplings are then RGE evolved 
\cite{Rothstein:1990qx} using the Mathematica package
SARAH \cite{Staub:2013tta}.

Using the same basis, the light neutrino mass matrix
can be taken as,
\begin{equation}
 m_{\nu} = U_l^* m_{\rm Diag} U_l^{\dagger};
 \quad U_l = P_e U_{PMNS} P_{\nu}; \quad m_{\rm Diag}
 = \text{Diag}(m_1, m_2, m_3).
\end{equation}
Here, $P_{\nu}$ refers to the Majorana phase matrix for 
light neutrinos and $P_e$ is the high energy phase matrix
of charged leptons. Because the low and high
energy regime neutrino masses follow the relation
$m_{\nu}(\text{high energy regime}) \simeq 1.3~ m_{\nu}
(\text{low energy regime})$ \cite{Giudice:2003jh,Antusch:2003kp},
the effect of RGE is nominal here.
Following these inputs the solutions for eight  
Majorana type Yukawa coupling matrices are calculated using the 
recipe given in section (\ref{sssec:3gen}).
For the RHN masses, using the mass matrix, 
\begin{equation}
M_R = fv_R,
\label{eq:rhn_mass}
\end{equation} 
eight different mass matrices can be obtained.
And correspondingly eight different mass spectra can be
derived for $M_1, M_2$, and $M_3$ by diagonalizing 
the mass matrix, $M_R$ and thus obtaining the 
diagonalising matrix $U_R$. In these mass expressions,
$v_R$ remains as a variable, which is the new physics scale.
These values are eventually used in the subsequent calculation
for AMM of muon and half-life of $0\nu\beta\beta$ taking TMM 
vertices of RHNs. In our earlier work \cite{Banerjee:2023aro} ( Appendix (B)),
we have shown the nature of $M_1, M_2$ and $M_3$ as a function of $v_R$.
\subsection{Transition Magnetic moment of RHNs in MLRSM}
\label{sec:nme}
The electromagnetic properties of neutrino
can be a window for new physics. For example neutrino magnetic
moments and neutrino masses are related to each other
since neutrino masses imply neutrino magnetic moments.
In MLRSM, four types of transition moments are possible, which are,
from processes; $\nu_i \rightarrow\nu_j\gamma$, $N_i \rightarrow N_j\gamma$, $\nu_i 
\rightarrow N_j\gamma$ and $N_i \rightarrow\nu_j\gamma$. They are 
expressed as \cite{Shrock:1982sc},
\begin{equation}
\begin{split}
&\mu_{\nu_i\nu_j} = -i \frac{3eg^2}{2^7\pi^2}(m_i + m_j)
\sum_{a=1}^{3}\left[\frac{\cos^2\zeta}{m_{W_1}^2}\epsilon_a^1 
+ \frac{\sin^2\zeta}{m_{W_2}^2}\epsilon_a^2 \right] 
\text{Im}[(U'_L)^*_{aj}(U'_L)_{ai}],
\\&\mu_{N_iN_j} = -i \frac{3eg^2}{2^7\pi^2}(M_i + M_j)
\sum_{a=1}^{3}\left[\frac{\sin^2\zeta}{m_{W_1}^2}\epsilon_a^1 
+ \frac{\cos^2\zeta}{m_{W_2}^2}\epsilon_a^2 \right] 
\text{Im}[(U'_R)^*_{aj}(U'_R)_{ai}],
\\&\mu_{\nu_iN_j}=i\frac{eg^2}{8\pi^2}\sin\zeta\cos\zeta
\sum_{a=1}^{3}m_{l_a} \text{Im}[e^{-i\phi}(U'_R)^*_{aj}(U'_L)_{ai}]
\sum_{b=1}^{2}\frac{(-1)^b}{m_{W_b}^2}\left(-1 + \epsilon_a^b 
\left\{\log\frac{1}{\epsilon_a^b} - \frac{9}{8}\right\} \right),
\\& \mu_{N_i\nu_j}= \mu_{\nu_iN_j}=\left( L\leftrightarrow R, 
\phi\rightarrow -\phi\right).
\end{split}
\label{eq:moments}
\end{equation}
In the above expressions for magnetic moments, the term 
$\epsilon_a^b$ is expressed as, $\epsilon_a^b =
\frac{m_{l_a}^2}{m_{W_b}^2}$. Here, $l$ stands for
the charged leptons.
In SM, the weak coupling constant $g$ is related with the Fermi
coupling constant $G_F$ by the relation, $\frac{G_F}{\sqrt{2}} =
\frac{g^2}{8M_W^2}$. Because of the extra gauge symmetry in 
MLRSM the $G_F$ can have additional corrections which will be 
reflected in the gauge coupling itself \cite{Czakon:2002vn}.
\section{Anomalous magnetic moment of Muon from magnetic moment of RHNs}
\label{sec:amm}
The anomalous magnetic moment of the muon, denoted by $a_{\mu}$
($a_\mu =(g-2)_\mu/2$)
(Ref. \cite{Jegerlehner:2009ry} for a review ), remains a 
significant topic in contemporary particle physics. Nearly two decades 
of pioneering efforts by the Brookhaven National Laboratory,  
Fermilab is now utilizing the same magnetic ring to measure $a_{\mu}$. The SM prediction for  
$a_{\mu}^{\text{SM}} = 116591810(43) \times 10^{-11}$ 
\cite{Aoyama:2020ynm}, whereas experimental measurements present a 
slightly higher value, $a_{\mu}^{\text{Expt}} = 116592061(41) 
\times 10^{-11}$ \cite{Muong-2:2021ojo}. This discrepancy suggests a 
potential contribution from physics beyond Standard Model (BSM) and is 
quantified as,
\begin{equation}
\Delta a_{\mu} = a_{\mu}^{\text{Expt}} - a_{\mu}^{\text{SM}} = 
(251 \pm 59) \times 10^{-11}.
\end{equation}
Also, recent lattice QCD (LQCD) calculations have made significant 
adjustments to the SM value for $a_\mu$ 
\cite{Blum:2019ugy, DellaMorte:2017dyu,  FermilabLatticeHPQCD:2023jof}. These adjustments, notably by the BMW
collaboration, have reduced the statistical significance of the
discrepancy from 4.2$\sigma$ to 1.6$\sigma$ \cite{DiLuzio:2021uty},
narrowing the gap as,
\begin{equation}
	\Delta a_{\mu} = (a_{\mu}^{\text{Expt}} - a_{\mu}^{\text{SM + LQCD}}) 
	= (107 \pm 72) \times 10^{-11}.
\end{equation} 
This small but highly sensitive gap, $\Delta a_{\mu}$, is a focal point 
in particle physics due to its potential to discover new physics. 
Precise measurements of $a_{\mu}$ serve as probes for BSM
fields and parameters.
The work by Queiroz \cite{Queiroz:2014zfa} explores all the possible
potential BSM field contributions, including scalar bosons, vector 
bosons, and leptoquarks, etc for $(g-2)_\mu$.

In this section, we study the role of the RHNs, through their 
magnetic moment vertex, in addressing the $(g-2)_\mu$ 
anomaly in MLRSM. Our study focuses on a Feynman 
diagram (Fig.\ref{fig:fd}) involving the RHN and the 
$W_R$ boson, with a unique vertex resembling a 
neutrino-gamma-neutrino interaction. 
In LRSM, the RHN masses are influenced by the interplay of 
type-I and type-II seesaw contributions.
Thus, the analysis provides to test the 
properties of Majorana-type Yukawa couplings and correspondingly
the dominance patterns of type-I and type-II seesaw. 
Before going into details,
it should be noted that the magnetic moment of light neutrinos 
is very well established and studied in numerous theoretical and 
experimental avenues. 
\subsection{$(g-2)_\mu$ due to the magnetic moment of RHNs in MLRSM}
\label{ssec:amm}
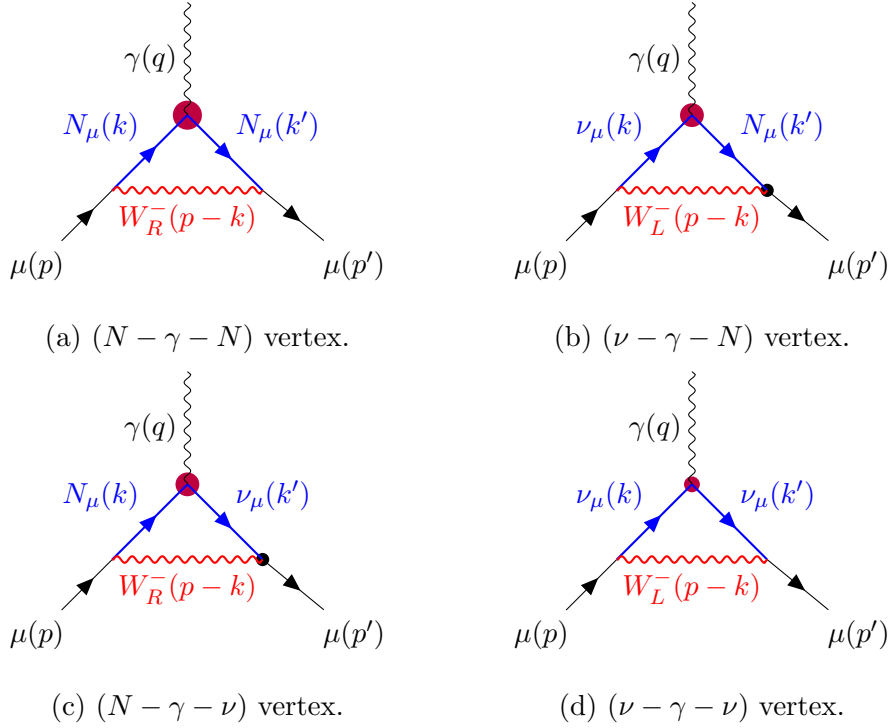
\begin{figure}[h]\centering
\begin{subfigure}[b]{0.40\textwidth}
\begin{tikzpicture}
\begin{feynman}
\vertex (a){\(\mu(p)\)};
\vertex[right=1cm of a](b);
\vertex[right=1cm of b](c);
\vertex[right=1cm of c](d);
\vertex[right=.67cm of d](e){\(\mu(p')\)};
\vertex[above=1cm of b](f);
\vertex[above=1cm of d](g);
\vertex[above=2cm of c](h);
\vertex[above=1.5cm of h](i);
\node[fill=purple, circle, minimum size=11pt, inner 
sep=0pt] (mid) at ($(h)!0!(h)$) {};
					
\diagram* {(a) -- [fermion](f) -- [fermion,thick,
blue,edge label=\(N_{\mu}(k)\)](h),(h) -- [fermion,
thick,blue,edge label=\(N_{\mu}(k')\)](g) -- 
[fermion](e),(f)  --[boson,thick,red,edge label'=
\(W^-_R(p - k)\)](g),(h) -- [boson,edge label=
\(\gamma(q)\)](i)};
\end{feynman}
\end{tikzpicture}
\caption{($N-\gamma-N$) vertex.}
\label{fig:1}
\end{subfigure}	
\begin{subfigure}[b]{0.40\textwidth}
\begin{tikzpicture}
\begin{feynman}
\vertex (a){\(\mu(p)\)};
\vertex[right=1cm of a](b);
\vertex[right=1cm of b](c);
\vertex[right=1cm of c](d);
\vertex[right=.67cm of d](e){\(\mu(p')\)};
\vertex[above=1cm of b](f);
\vertex[above=1cm of d](g);
\vertex[above=2cm of c](h);
\vertex[above=1.5cm of h](i);
\node[fill=purple, circle, minimum size=9pt, 
inner sep=0pt] (mid) at ($(h)!0!(h)$) {};
\node[fill=black, circle, minimum size=5pt,
inner sep=0pt] (mid) at ($(g)!0!(d)$) {};
					
\diagram* {(a) -- [fermion](f) -- [fermion,
thick,blue,edge label=\(\nu_{\mu}(k)\)](h),
(h) -- [fermion,thick,blue,edge label=
\(N_{\mu}(k')\)](g) -- [fermion](e),(f)  
--[boson,thick,red,edge label'=\(W^-_L(p - k)\)]
(g),(h) -- [boson,edge label=\(\gamma(q)\)](i)};
\end{feynman}
\end{tikzpicture}
\caption{($\nu-\gamma-N$) vertex.}
\label{fig:2}
\end{subfigure}	
\begin{subfigure}[b]{0.40\textwidth}
\begin{tikzpicture}
\begin{feynman}
\vertex (a){\(\mu(p)\)};
\vertex[right=1cm of a](b);
\vertex[right=1cm of b](c);
\vertex[right=1cm of c](d);
\vertex[right=.67cm of d](e){\(\mu(p')\)};
\vertex[above=1cm of b](f);
\vertex[above=1cm of d](g);
\vertex[above=2cm of c](h);
\vertex[above=1.5cm of h](i);
\node[fill=purple, circle, minimum size=9pt, 
inner sep=0pt] (mid) at ($(h)!0!(h)$) {};
\node[fill=black, circle, minimum size=5pt,
inner sep=0pt] (mid) at ($(g)!0!(d)$) {};
					
\diagram* {(a) -- [fermion](f) -- [fermion,
thick,blue,edge label=\(N_{\mu}(k)\)](h),(h)
-- [fermion,thick,blue,edge label=
\(\nu_{\mu}(k')\)](g) -- [fermion](e),
(f)  --[boson,thick,red,edge label'=
\(W^-_R(p - k)\)](g),(h) -- [boson,edge 
label=\(\gamma(q)\)](i)};
\end{feynman}
\end{tikzpicture}  
\caption{($N-\gamma-\nu$) vertex.}
\label{fig:3}
\end{subfigure}	
\begin{subfigure}[b]{0.40\textwidth}
\begin{tikzpicture}
\begin{feynman}
\vertex (a){\(\mu(p)\)};
\vertex[right=1cm of a](b);
\vertex[right=1cm of b](c);
\vertex[right=1cm of c](d);
\vertex[right=.67cm of d](e){\(\mu(p')\)};
\vertex[above=1cm of b](f);
\vertex[above=1cm of d](g);
\vertex[above=2cm of c](h);
\vertex[above=1.5cm of h](i);
\node[fill=purple, circle, minimum size=6pt,
inner sep=0pt] (mid) at ($(h)!0!(h)$) {};
					
\diagram* {(a) -- [fermion](f) -- 
[fermion,thick,blue,edge label=
\(\nu_{\mu}(k)\)](h),(h) -- 
[fermion,thick,blue,edge label=
\(\nu_{\mu}(k')\)](g) -- [fermion](e),
(f)  --[boson,thick,red,edge label'=
\(W^-_L(p - k)\)](g),(h) -- [boson,edge 
label=\(\gamma(q)\)](i)};
\end{feynman}
\end{tikzpicture}
\caption{($\nu-\gamma-\nu$) vertex.}
\label{fig:4}
\end{subfigure}	
\caption{Diagrams for radiative 
decay of muon consisting of RHNs, 
light neutrinos, and the left and right handed
gauge bosons. In each diagram, the
purple dots are the vertices that represent the
TMMs. Because of the $W_L-W_R$ mixing, in the 
calculation, the $W_L$ and $W_R$ are 
considered as the sum of mass eigenstates $W_1$ and $W_2$.} 
\label{fig:fd}
\end{figure}

In this section, we focus on the transition magnetic moment
vertices that appear in the calculation the 
muon AMM. In MLRSM, the RHNs contribute to the muon
AMM and our analysis shows that they can leave a significant 
contribution. The relevant terms in the 
Lagrangian are given by \cite{PhysRevD.16.2256, Petcov:1976ff, 
Boyarkin:2008zz},
\begin{equation}
\begin{split}
\mathcal{L} = \sum_{i\neq j} &[\mu_{N_iN_j} \bar{N_i}(k) 
\sigma^{\rho\nu}q_{\nu} N_j(k') + \mu_{N_i\nu_j} \bar{N_i}(k) 
\sigma^{\rho\nu}q_{\nu} \nu_j(k') \\& + \mu_{\nu_iN_j} 
\bar{\nu_i}(k) \sigma^{\rho\nu}q_{\nu} N_j(k') + \mu_{\nu_i\nu_j} 
\bar{\nu_i}(k) \sigma^{\rho\nu}q_{\nu} \nu_j(k')]A_{\rho}(q) + {\rm h.c} .
\end{split}
\end{equation} 
Each of the terms present in the Lagrangian corresponds to the diagrams
shown in Fig.(\ref{fig:fd}), consisting of right and left-chiral
muon neutrinos and $W_L$, $W_R$ gauge bosons in the loop.

The vertex function
corresponding to Fig.(\ref{fig:1}) is given by,
\begin{equation}
\begin{split}
\Gamma^{\rho} (q) = & \sum_{i\neq j}\mu_{N_iN_j}\int \frac{d^4 k}
{(2\pi)^4}\left[\frac{-ig}{2\sqrt{2}}\gamma^{\alpha}(1+\gamma_5)
\right] \left[\frac{i(\not k' + M_i)}{k^{'^{2}} - M_i^2 + i\epsilon}
\right] \sigma^{\rho\nu}q_{\nu} \left[\frac{i(\not k + M_j)}{k^{2} 
- M_j^2 + i\epsilon}\right]\\ & \times Z_{ij}^{RR} 
\left[\frac{-ig}{2\sqrt{2}}\gamma^{\beta}(1+\gamma_5)\right] 
\left[\frac{-ig_{\alpha\beta}}{(p-k)^2 - m_{W_R}^2 + i\epsilon} \right].
\end{split}
\label{eq:vf1}
\end{equation} 
In the above expression of vertex function, $\mu_{N_iN_j}$ 
is the TMM of RHN, $k$ and
$k' =(k+q)$ are momenta of incoming and outgoing RHNs
and $Z_{ij}^{RR}$ is the product of mixing matrix elements,
$Z_{ij}^{RR} =  ({U'}^{\dagger}_{R})_{i\mu} (U'_R)_{\mu i}
({U'}^{\dagger}_R)_{j\mu} (U'_R)_{\mu j}$ where $U'_R$ is 
the RHN mixing matrix that can be derived as shown
in Eq.(\ref{eq:6c6}). For the diagrams in figures (\ref{fig:2}),
(\ref{fig:3}), and (\ref{fig:4}) the respective products of
mixing matrix elements can be represented as,
\begin{equation}
\begin{split}
Z_{ij}^{LR} =  ({U'}^{\dagger}_{L})_{i\mu} (U'_L)_{\mu i} &
({U'}^{\dagger}_R)_{j\mu} (U'_R)_{\mu j},\hspace{15pt} 
Z_{ij}^{RL} =  ({U'}^{\dagger}_{R})_{i\mu} (U'_R)_{\mu i} 
({U'}^{\dagger}_L)_{j\mu} (U'_L)_{\mu j}, \\&Z_{ij}^{LL} =  
({U'}^{\dagger}_{L})_{i\mu} (U'_L)_{\mu i} ({U'}^{\dagger}_L)_{j\mu}
(U'_L)_{\mu j}.
\label{eq:mixing-Z}
\end{split}
\end{equation}
In Eq.(\ref{eq:vf1}), the propagator for $W_R$, can be decomposed
into the mass state  propagators of $W_1$ and $W_2$ using the 
relation in Eq.(\ref{eq:mixing}) (as shown in Appendix (\ref{app:d})). 
So, the vertex function in terms of mass states parameters is expressed as,
\begin{equation}
\begin{split}
\Gamma^{\rho} (q) = & \sum_{i\neq j}\mu_{N_iN_j}\int 
\frac{d^4 k}{(2\pi)^4}\left[\frac{-ig}{2\sqrt{2}}\gamma^{\alpha}
(1+\gamma_5)\right] \left[\frac{i(\not k' + M_i)}{k^{'^{2}} - 
M_i^2 + i\epsilon}\right] \sigma^{\rho\nu}q_{\nu} 
\left[\frac{i(\not k + M_j)}{k^{2} - M_j^2 + i\epsilon}\right]\\
& \times Z_{ij}^{RR} \left[\frac{-ig}{2\sqrt{2}}\gamma^{\beta}
(1+\gamma_5)\right] \left[-ig_{\alpha\beta}\left\{\frac{\sin^2\zeta}
{(p-k)^2 - m_{W_1}^2 + i\epsilon} + \frac{\cos^2\zeta}
{(p-k)^2 - m_{W_2}^2 + i\epsilon}\right\}\right].
\end{split}
\label{eq:vf2}
\end{equation} 
Now, using the Feynman parametrization technique \cite{PhysRev.76.769}
(the formula used here is shown in Appendix.(\ref{app:b}))
and the Clifford algebraic relations, the form factor for 
Fig.(\ref{fig:1}) can be extracted from the vertex function 
shown in Eq.(\ref{eq:vf2}). For example, for the form factor 
corresponding to $W_1$ part can be written as, 
\begin{equation}
\begin{split}
[F_2^{(a)}]_{W_1} \color{black}=& -
\frac{g^2 m_{\mu}\sin^2\zeta}{16e \pi^2}\sum_{i\neq j}
\mu_{N_iN_j}  Z_{ij}^{RR} \int_{0}^{1}dx \int_{0}^{1-x}dy \quad  
[4\slashed{q}(M_j + (M_i -M_j)(\frac{y}{2} - x)) \\& +
4M_iM_j + 16(y(1-2x)(m_{\mu}\slashed{q} + \frac{q^2}{2}) 
- x(1-x)q^2 + y^2m_{\mu}^2 +ym_{\mu}(M_i + M_j))] / \\ & 
[y(y-1)m_{\mu}^2 - x(1-x-y)q^2 + xM_i^2 + ym_{W_1}^2 + (1-x-y)M_j^2].
\end{split}
\label{eq:ammw1}
\end{equation}
Similarly, the form factor corresponding to the $W_2$ part can be obtained
by replacing the sine term with cosine and $m_{W_2}$ in the 
place of $m_{W_1}$, which is given by,
\begin{equation}
[F_{2}^{(a)}]_{W_2} = [F_2^{(a)}]_{W_1}(M_i,M_j,Z_{ij}^{RR},
m_{W_1} \rightarrow m_{W_2},\mu_{N_iN_j},\zeta \rightarrow \zeta 
+ \frac{\pi}{2},q^2).
\end{equation}
Similarly, for other diagrams the corresponding form factors
are calculated just by replacing the mass, magnetic moments 
and mixing angles correspondingly. For, Fig.(\ref{fig:2}) the
form factors are given by,
\begin{equation}
\begin{split}
&[F_{2}^{(b)}]_{W_1} =  [F_2^{(a)}]_{W_1}(M_i,M_j\rightarrow 
m_j,Z_{ij}^{RR} \rightarrow Z_{ij}^{LR}, m_{W_1},\mu_{\nu_iN_j},
\zeta \rightarrow \zeta + \frac{\pi}{2},q^2), \\&
[F_{2}^{(b)}]_{W_2} =  [F_2^{(a)}]_{W_1}(M_i,M_j\rightarrow m_j,
Z_{ij}^{RR} \rightarrow Z_{ij}^{LR}, m_{W_1}\rightarrow m_{W_2},
\mu_{\nu_iN_j},\zeta ,q^2).
\end{split}    	
\end{equation}
For, Fig.(\ref{fig:3}) those are,
\begin{equation}
\begin{split}
&[F_{2}^{(c)}]_{W_1} = 
[F_2^{(a)}]_{W_1}(M_i \rightarrow m_i,M_j,Z_{ij}^{RR} 
\rightarrow Z_{ij}^{RL}, m_{W_1},\mu_{N_i\nu_j},\zeta ,q^2), \\&
[F_{2}^{(c)}]_{W_2} = 
[F_2^{(a)}]_{W_1}(M_i \rightarrow m_i,M_j,Z_{ij}^{RR} 
\rightarrow Z_{ij}^{RL}, m_{W_1}\rightarrow m_{W_2},\mu_{N_i\nu_j},
\zeta \rightarrow \zeta + \frac{\pi}{2},q^2).
\end{split}    	
\end{equation}
and Fig.(\ref{fig:4}) driven form factors are,
\begin{equation}
\begin{split}
&[F_{2}^{(d)}]_{W_1} = 
[F_2^{(a)}]_{W_1}(M_i \rightarrow m_i,M_j \rightarrow m_j,Z_{ij}^{RR}
\rightarrow Z_{ij}^{LL}, m_{W_1},\mu_{\nu_i\nu_j},\zeta \rightarrow 
\zeta + \frac{\pi}{2},q^2), \\&
[F_{2}^{(d)}]_{W_2} = 
[F_2^{(a)}]_{W_1}(M_i \rightarrow m_i,M_j \rightarrow m_j,Z_{ij}^{RR} 
\rightarrow Z_{ij}^{LL}, m_{W_1}\rightarrow m_{W_2},\mu_{\nu_i\nu_j},
\zeta ,q^2).
\end{split}    	
\end{equation}
So, merging all the form factors results in  the complete expression
for $F_2^{total}$, given as
\begin{equation}
\begin{split}
F_2^{total} =  [F_2^{(a)}]_{W_1} + [F_{2}^{(a)}]_{W_2} + 
[F_{2}^{(b)}]_{W_1} + [F_{2}^{(b)}]_{W_2} +[F_{2}^{(c)}]_{W_1} 
+[F_{2}^{(c)}]_{W_2} + [F_{2}^{(d)}]_{W_1} + [F_{2}^{(d)}]_{W_2}.
\end{split}
\end{equation} \color{black}
	
The AMM for muon can be 
calculated from the term, $F_2^{total}$ at $q^2 = 0$. Now it can be
observed that, apart from other parameters of the theory,
the AMM for muon depends on the mass of the RHNs through
the term $F_2^{total}$. 
As the RHNs get mass from Majorana type Yukawa 
coupling matrices, $f$ given in Eq.(\ref{eq:nmlr}), it opens
up a window for investigating different seesaw dominance patterns 
of $f$ using the experimental results for the AMM for muon.
In the following subsection, we will discuss the results 
along with numerical values taken for all the parameters.
\subsection{Numerical analysis and results for muon anomalous
magnetic moment}
\label{sec:parameter_space}
The value of $F_2^{total}$ at $q^2=0$ gives the measure for the muon
AMM ($a_{\mu} = (g-2)_\mu/2 = F^{total}_{2}(q^2 = 0)$). In this section we present the details of the numerical analysis
and results related to the different seesaw dominance patterns 
of  Majorana coupling, $f$ using the experimental results for the AMM 
for muon. One can observe that $F_2^{total}$ is a multi-variable function,
which, apart from other parameters of the models, depends on the RHN 
masses, $M_1$, $M_2$ and $M_3$. For the RHN masses, using the mass 
matrix, $M_R = f v_R$, as discussed in section (\ref{ssec:dominance}),
eight different solutions of the mass matrix $M_R$ can be obtained.
And accordingly, upon diagonalization of $M_R$, eight different
combinations of $M_1$, $M_2$ and $M_3$ can be obtained depending 
upon the dominance of type-I and II seesaw in the light neutrino 
mass matrix. This feature provides a novel way to link high-energy
scale physics at low scale phenomena. Using these mass expressions,
only $v_R$ remains as a variable in the expression of $F_2^{total}$,
which is the new physics scale.

Apart from the dependence on the RHN masses, $F_2^{total}$ also 
depends on neutrino masses ($m_1$, $m_2$ and $m_3$), their mixing 
angles ($\theta_{12}$, $\theta_{13}$, and $\theta_{23}$), the Dirac
CP violating phase ($\delta_{CP}$), the Majorana phases ($\alpha$ 
and $\beta$), the gauge coupling constant ($g$), muon mass 
($m_{\mu}$), masses of the mass eigenstates of weak gauge bosons 
($m_{W_1}$, $m_{W_2}$), the mixing angle ($\zeta$) and the CP 
violating phase ($\phi$). The large number of variables are fixed
by their standard values and expressions. The lightest neutrino 
mass ($m_1$(NH)/$m_3$ (IH) is taken as $0.01$ eV. The rest of the
two neutrino masses and the mixing angles and the Dirac phase are 
calculated following the latest NuFIT data \cite{Esteban:2024eli}. 
The Majorana phases, $\alpha$ and $\beta$ are varied in the range
$0 -2\pi$ and the mass of muon is taken as $105.65$ MeV 
\cite{ParticleDataGroup:2020ssz}. The value of $m_{W_1}$ is approximately
equal to the mass of $W$ boson in SM, where the $m_{W_2}$ has 
a direct dependence on $v_R$ (shown in Eq.(\ref{eq:massofmass_st})).
In MLRSM, the value of the weak coupling constant $g_R=g_L=g$
gets a correction from radiative correction through Fermi coupling 
constant, $G_F$ \cite{Czakon:2002vn}. The corrected coupling 
constant value used in the calculation is $0.5386$.
	
The $W_L-W_R$ mixing angle has the dependence on $v_R$ as 
$\zeta= \frac{1}{2}\arctan\left(\frac{2k_1k_2}{v_R^2 - v_L^2}\right)$.
The experimental constraint is given by,  
$\zeta \leq 7.7 \times 10^{-4}$ \cite{ATLAS:2012ak, CMS:2012zv},
but its dependency on $v_R$ registers even smaller values with 
increasing $v_R$. So, for a sufficiently high scale of $v_R$, $\xi$ 
becomes very small.
The elements of the product of mixing matrix elements, 
like $Z_{ij}^{RR}$ can be calculated using equations (\ref{eq:6c6})
and (\ref{eq:mixing-Z}), Similarly the magnetic moments are
calculated using Eq.(\ref{eq:moments}) given in section (\ref{sec:nme}. 

\begin{figure}[h]
\includegraphics[height=6.5cm,width=11cm]{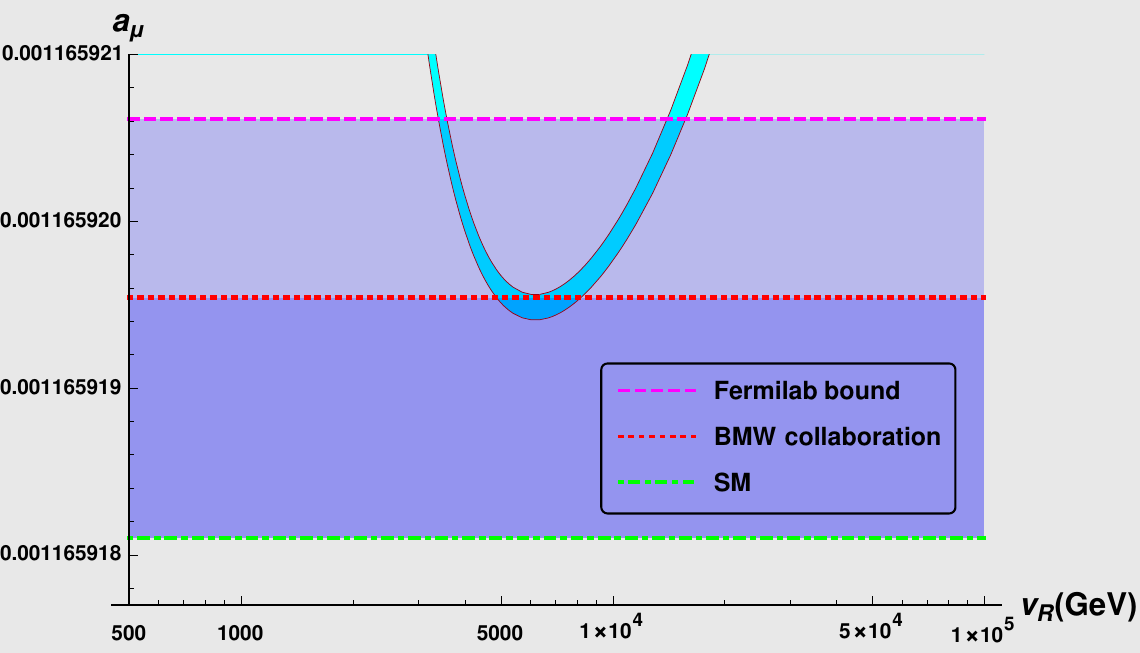}
\caption{The TMM contribution for the $a_{\mu}$ as a function 
of $v_R$ is shown in the above plot. The cyan colored region 
shows the contribution of $f$ for ($\pm++$) dominance patterns. The
variation of Majorana phases result in obtaining the cyan colored region.}
\label{fig:Majorana_+++}
\end{figure}

For the Majorana nature of neutrinos, the 
TMMs are responsible for the radiative
processes shown in figures (\ref{fig:1}),(\ref{fig:2}),
(\ref{fig:3}) and (\ref{fig:4}). The contributions 
from diagrams including the light neutrino mass, requires the
study of normal and inverted hierarchy in calculating the AMM 
of muon, In our analysis, the AMM of muon from magnetic moment 
vertices seem to be insensitive to the mass hierarchy of the 
light neutrinos. Also the study of different dominance patterns
in the RHN masses reveals that except from $+++$ and $-++$, the
rest of the patterns have diverging contributions to the AMM of
muon. In Fig.(\ref{fig:Majorana_+++}) and table (\ref{tab:valid_v_R_ranges}),
the combined contribution
of light neutrinos and RHNs, considering magnetic moment vertices,
are shown as a function of $v_R$.

\color{red}
\begin{table}[htb]
\centering
\setlength{\tabcolsep}{3pt} 
\renewcommand{\arraystretch}{1}
\begin{center}
\begin{tabular}{||c|c|c||} 
\hline
Dominance patterns & $a_{\mu}$ contribution ($\times 10^{-11}$) & $v_R$ range (TeV)  \\ [1.0ex] 
\hline
\hline
$+++$ & $116591938 - 116592061$ & $3.4 - 15$   \\ 
\hline
$-++$ & $116591938 - 116592061$ & $3.4 - 15$   \\
\hline
$+-+$ & $\gg$ Fermilab bound & $-$  \\
\hline
$--+$ & $\gg$ Fermilab bound & $-$  \\
\hline
$++-$ & $\gg$ Fermilab bound & $-$  \\
\hline
$-+-$ & $\gg$ Fermilab bound & $-$  \\
\hline
$+--$ & $\gg$ Fermilab bound & $-$  \\
\hline
$---$ & $\gg$ Fermilab bound & $-$  \\
\hline
\end{tabular}
\caption{Table shows the contribution to $a_\mu$ for all possible dominance patterns in $f$.  Only $(+++)$ and $(-++)$ cases have contributions within experimental predicted range for a feasible 
range of $v_R$. Whereas, the remaining patterns have divergent contributions for $a_\mu$.}
\label{tab:valid_v_R_ranges}
\end{center}
\end{table}
\color{black}

In Fig.(\ref{fig:Majorana_+++}), because of the variation of 
Majorana phases $\alpha$ and $\beta$ in the range $[0,2\pi]$,
the cyan colored region is observed which represents the 
contribution from TMMs of neutrinos with RHNs as a major 
contributor in MLRSM. As stated above only the  $+++$ and $-++$
case contribute to the AMM of muon within experimental sensitive
region, whereas other cases seesaw dominance result in diverging
contributions. The observed result shows that for the value of 
$v_R$ in the range, $\sim (3.4 \times 10^3 - 1.5 \times 10^4)$ GeV the 
diagrams in MLRSM considering the magnetic moment
vertices can saturate the experimental limit on $(g-2)_\mu$. 
This slightly distorted  parabolic shaped region signifies
interplay between the magnetic moments of 
neutrinos and the integral part as shown in Eq.(\ref{eq:ammw1}).
Initially, because of the smaller values of $W_2$ mass, the magnetic
moment $\mu_{N_iN_j}$, results in larger values. So, the plot attains
a monotonic decreasing nature up to $v_R \sim 6\times 10^3$ GeV. 
After that $\mu_{N_iN_j}$ saturates with increasing $v_R$ and becomes 
smaller than the integral part and the plot rises to the 
experimental average value at $v_R \sim 1.5 \times 10^4$ GeV.
Around $v_R \sim 6\times 10^3$ GeV,
the magnetic moment and integral part almost cancel each 
other and the sole contribution comes from the $W_1$ mass. 
	
This observation ensures that within the possible 
dominance patterns, the most stable Majorana type Yukawa 
coupling matrices, $(\pm ++)$ are the most viable ones, aligning with 
the studies performed in references \cite{Akhmedov:2006de} and \cite{Akhmedov:2006yp}. 
	
\section{Transition magnetic moment induced $0\nu\beta\beta$ decay}
\label{sec:TMM}
In this section we study $0\nu\beta\beta$ decay
in the presence of an external nonuniform magnetic field, where
the transition between neutrino and antineutrino can take place 
through the induced neutrino magnetic moment. 
Since neutrinos propagating in a nonuniform magnetic field may 
undergo the transition $\nu_\alpha \rightarrow \bar{\nu}_\alpha$, 
this process may mediate $0\nu\beta\beta$ decay. It was shown in references
\cite{Gozdz:2014pla, Gozdz:2014gna}, with a resonance like dependence
on the frequency of the external magnetic field, it will be possible
to enhance the rate of $0\nu\beta\beta$ decay by controlling the field.
Like in the previous section we consider the contribution of magnetic 
moment vertex of light neutrinos and RHNs in MLRSM to the decay process.
In this case the role of seesaw dominance patterns in RHN masses can be
studied, in forecasting the half-life of $0\nu\beta\beta$ decay. 
This study can be regarded as a complementary phenomenological analysis to $(g-2)_\mu$
in gauging the scale of $v_R$ as a new physics scale in $0\nu\beta\beta$ 
decay. Our study shows that for the range of $v_R$ in
$3.4 \times 10^3 - 1.5 \times 10^4$, obtained in the 
case of $(g-2)_\mu$, the half-life of $0\nu\beta\beta$ decay
falls in a range which is experimentally inaccessible in present and future
searches of $0\nu\beta\beta$ decay. So the diagrams including weak interaction 
vertex remain as potential contributors to $0\nu\beta\beta$ decay in MLRSM 
as we have shown in our previous study \cite{Banerjee:2023aro}.

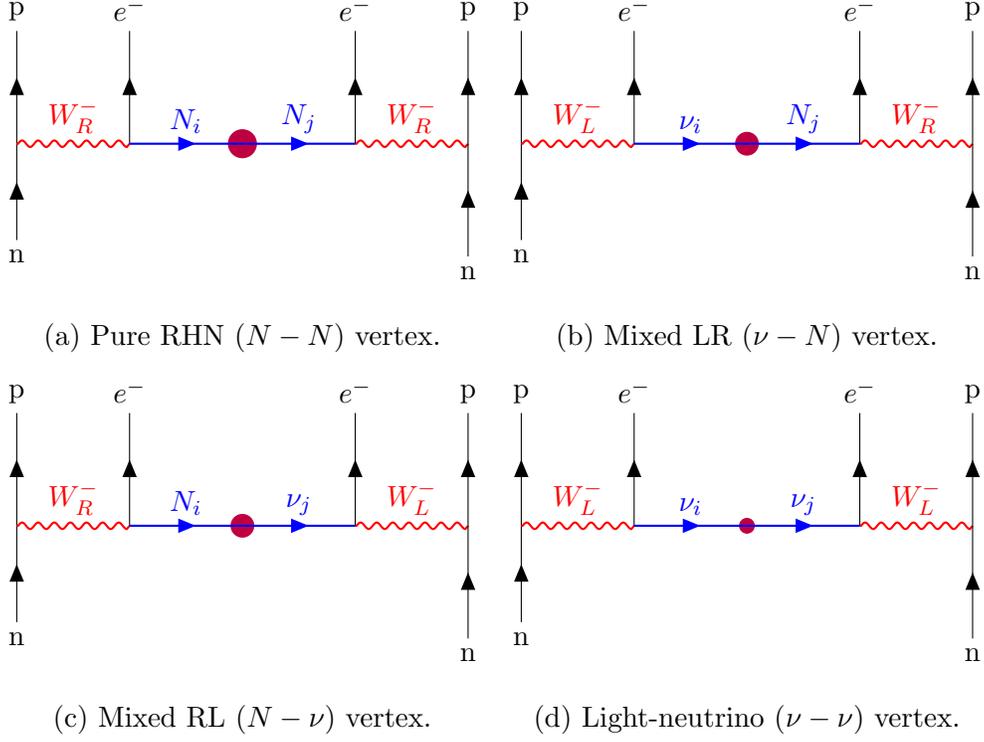
\begin{figure}[h]\centering
\begin{subfigure}[b]{0.4\textwidth}
\begin{tikzpicture}
\begin{feynman}
\vertex (a) {n};
\vertex [above=of a](b) ;
\vertex [above=of b](c) {p};
\vertex [right=of b](d) ;
\vertex [right=of d](e) ;
\vertex [right=of e](f) ;
\vertex [right=of f](g) ;
\vertex [below=of g](h) {n};
\vertex [above=of g](i) {p};
\vertex [above=of d](j) {$e^-$};
\vertex [above=of f](l) {$e^-$};
\node[fill=purple, circle, minimum size=11pt, 
inner sep=0pt] (mid) at ($(e)!0!(f)$) {};
					
\diagram{(a) --[fermion](b);
(b) --[fermion](c);
(b) --[boson, edge label=$W^-_{R}$, red, thick](d);
(f) --[boson, edge label=$W^-_{R}$, red, thick](g);
(h) --[fermion](g);
(g) --[fermion](i);
(d) --[fermion](j);
(f) --[fermion](l);
(d) --[fermion, edge label=$N_i$, blue, thick](e);
(e) --[fermion, edge label=$N_j$, blue, thick](f);
};
\end{feynman}
\end{tikzpicture}
\caption{Pure RHN ($N-N$) vertex.}
\label{fig:1a}
\end{subfigure}	
\begin{subfigure}[b]{0.4\textwidth}
\begin{tikzpicture}
\begin{feynman}
\vertex (a) {n};
\vertex [above=of a](b) ;
\vertex [above=of b](c) {p};
\vertex [right=of b](d) ;
\vertex [right=of d](e) ;
\vertex [right=of e](f) ;
\vertex [right=of f](g) ;
\vertex [below=of g](h) {n};
\vertex [above=of g](i) {p};
\vertex [above=of d](j) {$e^-$};
\vertex [above=of f](l) {$e^-$};
\node[fill=purple, circle, minimum size=9pt, 
inner sep=0pt] (mid) at ($(e)!0!(f)$) {};
					
\diagram{(a) --[fermion](b);
(b) --[fermion](c);
(b) --[boson, edge label=$W^-_{L}$, red, thick](d);
(f) --[boson, edge label=$W^-_{R}$, red, thick](g);
(h) --[fermion](g);
(g) --[fermion](i);
(d) --[fermion](j);
(f) --[fermion](l);
(d) --[fermion, edge label=$\nu_i$, blue, thick](e);
(e) --[fermion, edge label=$N_j$, blue, thick](f);
};
\end{feynman}
\end{tikzpicture}
\caption{Mixed LR ($\nu-N$) vertex.}
\label{fig:2a}
\end{subfigure}	
\begin{subfigure}[b]{0.4\textwidth}
\begin{tikzpicture}
\begin{feynman}
\vertex (a) {n};
\vertex [above=of a](b) ;
\vertex [above=of b](c) {p};
\vertex [right=of b](d) ;
\vertex [right=of d](e) ;
\vertex [right=of e](f) ;
\vertex [right=of f](g) ;
\vertex [below=of g](h) {n};
\vertex [above=of g](i) {p};
\vertex [above=of d](j) {$e^-$};
\vertex [above=of f](l) {$e^-$};
\node[fill=purple, circle, minimum size=9pt, 
inner sep=0pt] (mid) at ($(e)!0!(f)$) {};
					
\diagram{(a) --[fermion](b);
(b) --[fermion](c);
(b) --[boson, edge label=$W^-_{R}$, red, thick](d);
(f) --[boson, edge label=$W^-_{L}$, red, thick](g);
(h) --[fermion](g);
(g) --[fermion](i);
(d) --[fermion](j);
(f) --[fermion](l);
(d) --[fermion, edge label=$N_i$, blue, thick](e);
(e) --[fermion, edge label=$\nu_j$, blue, thick](f);
};
\end{feynman}
\end{tikzpicture}  
\caption{Mixed RL ($N-\nu$) vertex.}
\label{fig:3a}
\end{subfigure}	
\begin{subfigure}[b]{0.4\textwidth}
\begin{tikzpicture}
\begin{feynman}
\vertex (a) {n};
\vertex [above=of a](b) ;
\vertex [above=of b](c) {p};
\vertex [right=of b](d) ;
\vertex [right=of d](e) ;
\vertex [right=of e](f) ;
\vertex [right=of f](g) ;
\vertex [below=of g](h) {n};
\vertex [above=of g](i) {p};
\vertex [above=of d](j) {$e^-$};
\vertex [above=of f](l) {$e^-$};
\node[fill=purple, circle, minimum size=6pt,
inner sep=0pt] (mid) at ($(e)!0!(f)$) {};
					
\diagram{(a) --[fermion](b);
(b) --[fermion](c);
(b) --[boson, edge label=$W^-_{L}$, red, thick](d);
(f) --[boson, edge label=$W^-_{L}$, red, thick](g);
(h) --[fermion](g);
(g) --[fermion](i);
(d) --[fermion](j);
(f) --[fermion](l);
(d) --[fermion, edge label=$\nu_i$, blue, thick](e);
(e) --[fermion, edge label=$\nu_j$, blue, thick](f);
};
\end{feynman}
\end{tikzpicture}
\caption{Light-neutrino ($\nu-\nu$) vertex.}
\label{fig:4a}
\end{subfigure}	
\caption{Diagrams for $0\nu\beta\beta$ decay where 
the TMMs of RHN, light neutrino, and mixed 
cases are shown. The magenta dots in the diagrams 
represent TMM vertices.}
\label{fig:transition}
\end{figure}

The very basic diagram for neutrinoless 
double beta decay consists of a neutrino annihilation in the middle
of two beta decay events, which is the clue for the Majorana nature
of the neutrino. A different situation can also take place, that 
is the transition of the neutrino state to the anti-neutrino state
via the magnetic moment of the neutrino as shown in 
Fig.(\ref{fig:transition}). This case is also valid for the 
RHNs in MLRSM. This scenario offers to study the seesaw dominance 
patterns through the Majorana type Yukawa coupling inclusion in 
the RHN mass (Eq.\ref{eq:rhn_mass}), in the presence
of a finite magnetic field. As shown in section.(\ref{sec:nme}), 
there are four transition moments for which four corresponding
diagrams are possible as shown in Fig.(\ref{fig:transition}). 

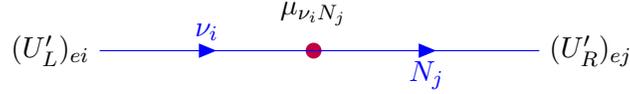
\begin{figure}[h]
\begin{tikzpicture}
\begin{feynman}
\vertex (a) {$(U'_L)_{ei}$};
\vertex [right=3.5cm of a](b) ;
\vertex [right=3cm of b](c) {$(U'_R)_{ej}$};
\vertex [above=0.2cm of b](d) {$\mu_{\nu_i N_j}$};
\node[fill=purple, circle, minimum size=6pt,inner sep=0pt] 
(mid) at ($(b)!0!(b)$) {};
\diagram{(a) --[fermion, edge label=$\nu_i$, blue](b);
(b) --[fermion, edge label'=$N_j$, blue](c); };
\end{feynman}
\end{tikzpicture}
\caption{The internal transition magnetic moment line of 
$0\nu\beta\beta$ decay process for Fig(\ref{fig:2a}).}
\label{fig:tmm_line}
\end{figure}

Contributions from this type of TMM induced diagrams come 
from the internal line present in each diagram. For example,
the internal line of Fig.(\ref{fig:2a}) is shown in Fig.(\ref{fig:tmm_line}).
The lepton number violating contribution coming from this transition
line is proportional to the product of propagators, mixing matrices
present at each end of the propagators and the magnetic moment of 
neutrino in the mass basis. The quantity that represents this TMM
induced interaction is expressed by the term $\chi_{LR}$. So, for 
Fig.(\ref{fig:2a}), the $\chi_{LR}$ term is defined as 
\cite{Gozdz:2014pla, Gozdz:2014gna},
\begin{equation}
\begin{split}
\chi_{LR} = \sum_{\alpha,\beta}\mu_{\alpha\beta}^{LR} 
\left(\sum_{i=1,2,3}(U'_L)_{ei} \frac{\slashed{p} + m_i}
{p^2 - m_i^2}(U'_L)_{\alpha i}^* \right)\left(\sum_{j=1,2,3}
(U'_R)_{\beta j}^* \frac{\slashed{p} + M_j}{p^2 - M_j^2}
(U'_R)_{ej}^*\right).
\end{split}
\end{equation}
Here, $U'_L$ and $U'_R$ are the mixing matrices associated with 
the left and right-handed neutrino sectors, respectively, and
$m_i$, $M_j$ denote their corresponding mass eigenvalues. 
And $\mu^{LR}_{\alpha\beta}$ is the magnetic moment in flavor basis 
which is connected to magnetic moment in mass basis by the 
relation,
\begin{equation}
 \mu_{\nu_iN_j}=\sum_{\alpha,\beta}(U'_L)_{\alpha i}^*
\mu_{\alpha\beta}^{LR}(U'_R)_{\beta j} .
\end{equation}

Similarly the expressions for $\chi_{RR}, \chi_{RL}$ and 
$\chi_{LL}$ for the diagrams Fig.(\ref{fig:1a}), (\ref{fig:3a}) and 
(\ref{fig:4a}) respectively, are expressed as,
\begin{equation}
\begin{split}
&\chi_{RR} = \sum_{\alpha,\beta}\mu_{\alpha\beta}^{RR} 
\left(\sum_{i=1,2,3}(U'_R)_{ei} \frac{\slashed{p} + M_i}
{p^2 - M_i^2}(U'_R)_{\alpha i}^* \right)\left(\sum_{j=1,2,3}
(U'_R)_{\beta j}^* \frac{\slashed{p} + M_j}{p^2 - M_j^2}
(U'_R)_{ej}^*\right),  \\&
\chi_{RL} = \sum_{\alpha,\beta}\mu_{\alpha\beta}^{RL}
\left(\sum_{i=1,2,3}(U'_R)_{ei} \frac{\slashed{p} + M_i}
{p^2 - M_i^2}(U'_R)_{\alpha i}^* \right)\left(\sum_{j=1,2,3}
(U'_L)_{\beta j}^* \frac{\slashed{p} + m_j}{p^2 - m_j^2}
(U'_L)_{ej}^*\right), \\&
\chi_{LL} = \sum_{\alpha,\beta}\mu_{\alpha\beta}^{LL} 
\left(\sum_{i=1,2,3}(U'_L)_{ei} \frac{\slashed{p} + m_i}
{p^2 - m_i^2}(U'_L)_{\alpha i}^* \right)\left(\sum_{j=1,2,3}
(U'_L)_{\beta j}^* \frac{\slashed{p} + m_j}{p^2 - m_j^2}
(U'_L)_{ej}^*\right).
\end{split}
\end{equation}                
In the expressions, the magnetic moments, $\mu_{\alpha\beta}^{RR}, \mu_{\alpha\beta}^{RL}, \mu_{\alpha\beta}^{LL}$
are in the flavor basis which is related to the magnetic moments 
of the neutrino mass-basis by the following relations,
\begin{equation}
\begin{split}
&\mu_{N_iN_j}=\sum_{\alpha,\beta}(U'_R)_{\alpha i}^*
\mu_{\alpha\beta}^{RR}(U'_R)_{\beta j}, \hspace{10pt} 
\mu_{N_i\nu_j}=\sum_{\alpha,\beta}(U'_R)_{\alpha i}^*
\mu_{\alpha\beta}^{RL}(U'_L)_{\beta j},
\hspace{10pt} \mu_{\nu_i\nu_j}=\sum_{\alpha,\beta}
(U'_L)_{\alpha i}^*\mu_{\alpha\beta}^{LL}(U'_L)_{\beta j}.
\end{split}
\end{equation}
  
So, from Fig.(\ref{fig:transition}), the amplitudes 
for the diagrams can be written as, 
\begin{equation}
\begin{split}	
&\mathcal{A}_{TMM}^{RR} \propto \left(\frac{m_{W_L}}{m_{W_R}}
\right)^4 \lvert B\chi_{RR} \lvert^2, \hspace{15pt}
\mathcal{A}_{TMM}^{LR} \propto \left(\frac{m_{W_L}}{m_{W_R}}
\right)^4 \lvert B\chi_{LR} \lvert^2,  \\&
\mathcal{A}_{TMM}^{RL} \propto \left(\frac{m_{W_L}}{m_{W_R}}
\right)^4 \lvert B\chi_{RL} \lvert^2, \hspace{15pt}
\mathcal{A}_{TMM}^{LL} \propto \lvert B\chi_{LL} \lvert^2.
\end{split}
\end{equation}
Here, $B$ is the external magnetic field and the dimensionless
particle physics terms relevant to these amplitudes are,
\begin{equation}
\begin{split}
&\eta_{TMM}^{RR} = \left(\frac{m_{W_L}}{m_{W_R}}\right)^4 
\lvert B\chi_{RR} \lvert^2 m_p^2, \hspace{15pt}
\eta_{TMM}^{LR} = \left(\frac{m_{W_L}}{m_{W_R}}\right)^4 
\lvert B\chi_{LR} \lvert^2 m_p^2, \\&
\eta_{TMM}^{RL} = \left(\frac{m_{W_L}}{m_{W_R}}\right)^4 
\lvert B\chi_{RL} \lvert^2 m_p^2, \hspace{15pt}
\eta_{TMM}^{LL} =  \lvert B\chi_{LL} \lvert^2 m_e^2.
\end{split}
\end{equation}
Here $m_p$ and $m_e$ are the masses of proton and electron. 
Using these $\eta_{TMM}$ terms, the lifetime for $0\nu\beta\beta$ 
decay can be calculated as,
\begin{equation}
\left[T_{1/2}^{0\nu\beta\beta}\right]^{-1} = G_{0\nu} \left(\lvert 
\boldsymbol{M_{\nu}^{0\nu}} \lvert^2 \lvert \eta_{TMM}^{LL}\lvert^2
+ \lvert \boldsymbol{M_N^{0\nu}} \lvert^2 \lvert \eta_{TMM}^{LR} +
\eta_{TMM}^{RL}+ \eta_{TMM}^{RR} \lvert^2\right).
\label{eq:halflife}
\end{equation}
The nuclear matrix elements (NME) $\boldsymbol{M_{\nu}^{0\nu}}$
and $\boldsymbol{M_N^{0\nu}}$, take care of the hadronic part 
involved in the possible diagrams shown in Fig.(\ref{fig:transition})
and $G_{0\nu}$ is the phase space factor. 
	
The unique feature of TMM-induced decay is that, 
the decay width can be modulated by changing the intensity 
of the magnetic field. So, a resonance criteria can be 
created for the direct observation of $0\nu\beta\beta$ decay.  
	
\subsection{Results for different magnetic field and dominance patterns}
\label{ssec:onbb_TMM_result}
For $0\nu\beta\beta$ decay process, the half-life in
Eq.(\ref{eq:halflife}) is related with phase factor, 
NMEs and the dimensionless particle physics parameters. 
The values for the NMEs and phase space factors that are 
used in the calculation are 
shown in table (\ref{tab:npv}).
\begin{table}[htb]
\centering
\setlength{\tabcolsep}{8pt} 
\renewcommand{\arraystretch}{1.2}
\begin{center}
\begin{tabular}{||c|c|c|c||} 
\hline
Isotope & $G_{0\nu}(yr^{-1})$ & $\mathbf{M}_{\nu}^{0\nu}$ & 
$\mathbf{M}_{N}^{0\nu}$  \\ [1.0ex] 
\hline\hline
Ge-76 & $5.77\times10^{-15}$ & 2.58-6.64 & 233-412  \\ 
\hline
Xe-136 & $3.56\times10^{-14}$ & 1.57-3.85 & 164-172  \\ 
\hline
\end{tabular}
\caption{Values taken for the phase factor \cite{Kotila:2012zza} 
and nuclear matrix elements \cite{Dueck:2011hu, Faessler:2011rv,
Faessler:2011qw} used in calculation.}
\label{tab:npv}
\end{center}
\end{table}
The $\eta_{TMM}$ terms incorporate all possible parameters of the MLRSM,
along with the masses of the electron ($m_e$) and proton ($m_p$).
In this analysis, the MLRSM parameters are adopted as specified 
in Section.(\ref{sec:parameter_space}), with $m_e = 0.511$ eV 
and $m_p = 931.49$ MeV. For $0\nu\beta\beta$ decay the last updated
bound for half-life is $10^{26}$ Year \cite{KamLAND-Zen:2022tow,
Shtembari:2022pfs} and expected bound in near future is $10^{28}$ years 
\cite{nEXO:2017nam, LEGEND:2017cdu, Armengaud:2019loe}. 

A crucial parameter is the external magnetic field $B$,
which significantly influences the lifetime of $0\nu\beta\beta$ 
decay. As on the Earth's surface, magnetic field
remains approximately constant at $B_0 = 30$ $\mu$T 
\cite{Dmitriev_2016}, we explore the variation of the external magnetic 
field parameter $B$ as multiples of $B_0$. A systematic study of these 
variations allows us to probe their impact on the phenomenology of 
$0\nu\beta\beta$ decay, particularly in the presence of right-handed 
currents. We take the viable range of $v_R$ for explaining the $(g-2)_\mu$, i.e., $v_R \sim 3.4 \times 10^3 - 1.5 \times 10^4$  GeV, in 
estimating the half-life of  
$0\nu\beta\beta$ decay. Within this range, the predicted half-life
is $\mathcal{O}(10^{160})$ years at Earth's magnetic field strength.
Increasing the magnetic field up to a million times that of Earth 
reduces the half-life to $\mathcal{O}(10^{120})$ years. Despite 
this reduction, both values remain far beyond the reach of current
and future experiments, indicating that the contribution from TMM
vertices is highly suppressed. This reinforces the conclusion that
weak interaction vertices play a more dominant role in 
$0\nu\beta\beta$ decay. 
	
\color{black}	
\section{Conclusion}
\label{sec:discussion}
Although in LRSM, unlike Dirac type coupling,  Majorana type 
coupling can not be constrained experimentally, nevertheless,
the seesaw formula including both type-I and type-II contributions 
can be employed to reconstruct the Majorana
coupling matrix taking certain quantities as input parameters
\cite{Akhmedov:2006de}. The study can provide some insight in to
the underlying theory at the seesaw scale. Taking experimentally
observable data for
light neutrino masses and mixing as input and assuming up-quark 
type Dirac Yukawa coupling for the neutrinos,
the Majorana coupling (and hence the masses of the right-handed
neutrinos) can be expressed numerically as function of the left-right
symmetry breaking scale, $v_R$ \cite{Hosteins:2006ja}. Depending 
upon dominance of type-I and type-II
seesaw terms and interplay among them, total eight solutions for 
the Majorana coupling are obtained. We have studied the signatures 
of the eight solutions taking the electromagnetic property of both
light neutrinos and RHNs. By taking the transition magnetic moment
(for their Majorana nature in MLRSM) vertex in second order processes
like $\mu \rightarrow \mu \gamma$ and $0\nu\beta\beta$ decay process we
have studied the role of eight solutions for the Majorana coupling matrix
through the dependence of the latter processes on RHN masses, taking 
$v_R$ as the new physics scale. 
Out of the eight solutions, only $+++$ and $-++$ solutions only contribute
to the $(g-2)_\mu$ in the experimental predicted range. The range
of $v_R$ in this case is found to be $3.4 \times 10^3 - 1.5 \times 10^4$  GeV.
To check the compatibility of the prediction, thus obtained, in predicting
the half-life of $0\nu\beta\beta$ decay we find that for the same range of $v_R$
the half-life of $0\nu\beta\beta$  decay falls in a range which is far beyond the 
reach of current and future searches of $0\nu\beta\beta$ decay.
 This signals that the usual mass mechanism of $0\nu\beta\beta$ decay
should be the dominant. Being a higher
order process, the magnetic moment channel seems to be
subdominant to the mass channel of $0\nu\beta\beta$ decay.
\section{Acknowledgements}
We extend our heartfelt gratitude to Prof. Evgeny Akhmedov 
for sharing invaluable insights and enriching discussions
regarding the work. V.B would like to thank  Prof. Michael
Schmidt for constructive suggestions at the PPC 2024 conference
at IIT Hyderabad and also to the Ministry of Human Resource
Development (MHRD), Government of India, for their generous 
support.
\section{Appendix}
\subsection{Field decomposition of the gauge fields}
\label{app:d}
In the language of Quantum Field Theory, the propagator of a
field is a two-point correlation function. Starting with simple
scalar field $\phi$, its decomposition can be defined as,
\begin{equation}
 \phi(x) = \phi^+(x) + \phi^-(x).
\end{equation}
This $\phi^-(x)$ and $\phi^+(x)$ contain the creation and 
annihilation operators respectively. So, they follow the conditions, 
\begin{equation}
\phi^+(x) |0\rangle = 0, \hspace{10pt} \langle 0|\phi^-(x) = 0.
\end{equation}
The propagator of this field is \cite{Peskin:1995ev},
\begin{equation}
\langle 0|T\{\phi(x)\phi(y)\}|0\rangle_{(x^0 > y^0)} = [\phi^+(x), 
\phi^-(y)] = D(x - y).
\end{equation}
Now, from Eq.(\ref{eq:mixing}), the $W_R$ state can be 
expressed as, $W_R = -W_1\sin\zeta  + W_2\cos\zeta $,
by taking $\phi$ as zero. So, the $W_R$ field appears
as the linear combination of $W_1$ and $W_2$ mass states.
Now, performing the similar treatment like scalar field,
the two-point correlation function for $W_R$ stands as,
\begin{equation}
\begin{split}
\langle 0|T\{W_R(x)W_R(y)\}|0\rangle_{(x^0 > y^0)} &= 
[W_R^+(x), W_R^-(y)] \\& = [W_1^+(x), W_1^-(y)]\sin^2\zeta 
+ [W_2^+(x), W_2^-(y)]\cos^2\zeta.
\end{split}
\end{equation}
This relation is used in Eq.(\ref{eq:vf1}) to express 
the $W_R$ propagator in terms of mass states propagators
as shown in Eq.(\ref{eq:vf2}). Similarly, following the
same analogy one can write down the $W_L$ propagator in
terms of mass state 
propagators also. 
	
\subsection{Field theoretic description of $a_{\mu}$}
\label{sec:peskin}
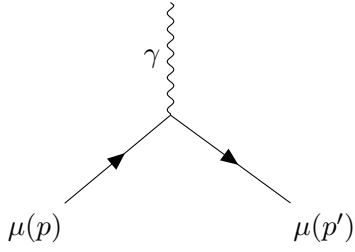
\begin{figure}[h]\centering
\begin{tikzpicture}
\begin{feynman}
\vertex (a){\(\mu(p)\)};
\vertex[right=1.8cm of a](b);
\vertex[above=1.5cm of b](c);
\vertex[right=1.5cm of b](d){\(\mu(p')\)};
\vertex[above=1.5cm of c](e);
\diagram* {(a) -- [fermion](c),(c) -- [fermion](d),(c) -- 
[boson,edge label = \(\gamma\)](e)};
\end{feynman}
\end{tikzpicture}
\caption{The tree level Feynman diagram of muon electromagnetic
interaction.}
\label{fig:simplest}
\end{figure}

The quantum field theoretical explanation for $a_{\mu}$, allows one
to explore the phenomenon with a large number of possibilities. 
In general, the tree level Feynman diagram has one incoming and 
one outgoing muon with a gamma interaction as in 
Fig.(\ref{fig:simplest}).
The amplitude for this diagram is, 
\begin{equation}
iM^{\mu} = -ie\bar{u}(p^{'}) \gamma^{\mu}u(p).
\end{equation}
	
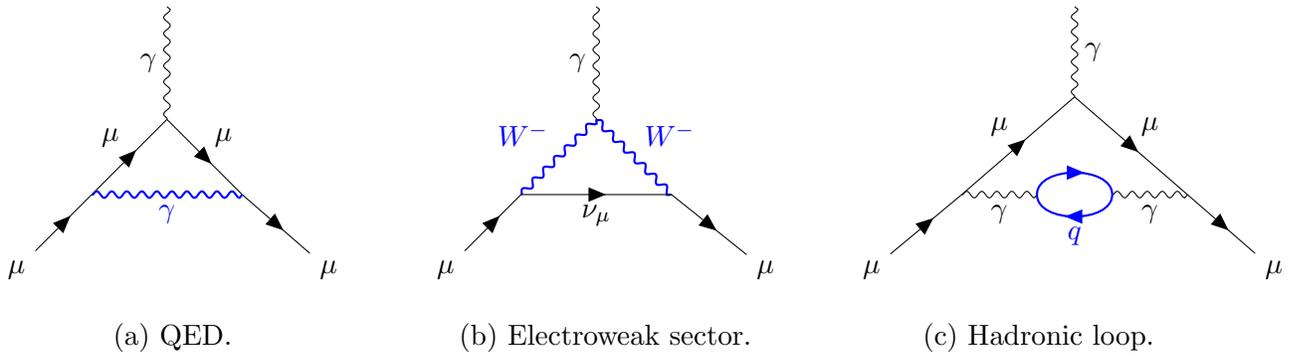
\begin{figure}[h]
\centering
\begin{subfigure}[b]{0.30\textwidth}
\begin{tikzpicture}
\begin{feynman}
\vertex (a){\(\mu\)};
\vertex[right=1cm of a](b);
\vertex[right=1cm of b](c);
\vertex[right=1cm of c](d);
\vertex[right=0.9cm of d](e){\(\mu\)};
\vertex[above=1cm of b](f);
\vertex[above=1cm of d](g);
\vertex[above=2cm of c](h);
\vertex[above=1.5cm of h](i);\diagram* {(a) -- [fermion](f) --
[fermion,edge label=\(\mu\)](h),(h) -- 
[fermion,edge label=\(\mu\)](g) -- [fermion](e),(f) 
--[boson,thick,blue,edge label'=\(\gamma\)](g),(h) 
-- [boson,edge label=\(\gamma\)](i)};
\end{feynman}
\end{tikzpicture}
\caption{QED.}
\end{subfigure}
\hfill
\begin{subfigure}[b]{0.30\textwidth}
\centering
\begin{tikzpicture}
\begin{feynman}
\vertex (a){\(\mu\)};
\vertex[right=1cm of a](b);
\vertex[right=1cm of b](c);
\vertex[right=1cm of c](d);
\vertex[right=1cm of d](e){\(\mu\)};
\vertex[above=1cm of b](f);
\vertex[above=1cm of d](g);
\vertex[above=2cm of c](h);
\vertex[above=1.5cm of h](i);
\diagram* {(a) -- [fermion](f) --
[boson,thick,blue,edge label=\(W^-\)](h),(h) 
-- [boson,thick,blue,edge label=\(W^-\)](g) --
[fermion](e),(f)  --[fermion,edge label'=
\(\nu_{\mu}\)](g),(h) -- [boson,edge label=
\(\gamma\)](i)};
\end{feynman}
\end{tikzpicture}
\caption{Electroweak sector.}
\end{subfigure}
\hfill
\begin{subfigure}[b]{0.30\textwidth}
\centering
\begin{tikzpicture}
\begin{feynman}
\vertex (a){\(\mu\)};
					\vertex[right=1.2cm of a](b);
					\vertex[right=1cm of b](c);
					\vertex[right=0.5cm of c](d);
					\vertex[right=0.5cm of d](e);
					\vertex[right=1cm of e](f);
					\vertex[right=0.9cm of f](g){\(\mu\)};
					\vertex[above=1cm of b](i);
					\vertex[above=1cm of c](j);
					\vertex[above=1cm of e](k);
					\vertex[above=1cm of f](l);
					\vertex[above=2.3cm of d](m);
					\vertex[above=1.2cm of m](n);
					\vertex[above=0.3cm of d](p);
					\vertex[above=2.7cm of p](o);
					\diagram* {(a) -- [fermion](i) -- 
					[fermion,edge label=\(\mu\)](m),(m)
					-- [fermion,edge label=\(\mu\)](l) 
					-- [fermion](g),(i)  --[boson,edge 
					label'=\(\gamma\)](j),(k)  --
					[boson,edge label'=\(\gamma\)](l),(m)
					-- [boson,edge label'=\(\gamma\)](n),(j)
					-- [fermion,thick,blue,out=90, in=90,
					looseness=1.0](k),(k) -- 
					[fermion,thick,blue,out=270, 
					in=270, looseness=1.0, edge label=\(q\)](j)};
				\end{feynman}
			\end{tikzpicture}
			\caption{Hadronic loop.}
		\end{subfigure}
\caption{Examples of 1st-order loop diagrams for muon $(g - 2)$,
with the fields present in SM.}
\label{fig:fdm}
\end{figure}

However, this simple diagram can have various higher order loops. 
In general, depending on the nature of interactions they can be 
categorized into three types as shown in Fig.(\ref{fig:fdm}). 
The first one is for QED contribution, second and third are 
the weak \cite{Czarnecki:2002nt, Bars:1972pe} and hadronic
contributions respectively. So, the value attained for muon 
anomaly ($a_{\mu}$) in the SM can be separated into three
components,
\begin{equation*}
		a^{\text{SM}}_{\mu} = 
		a_{\mu}^{\text{QED}} + a_{\mu}^{\text{Electroweak}} 
		+ a_{\mu}^{\text{Hadronic}}.
	\end{equation*} 
Problem arises when the experimentally measured value for 
$a_{\mu}$ becomes a bit (of order $10^{-11}$) higher than the
SM calculation. This very small increased value is a matter of 
concern because it can lead to a number of important consequences.
It can act as a probe for detecting the so called BSM fields and 
draw constraints on additional parameters.
	
For possible 1-loop Feynman diagrams of a muon with incoming 
momentum $p$ and outgoing momentum $p'$ interacting with photons 
and other fermions and bosons, the amplitude has a compact form like,
\begin{equation}
iM^{\mu} = -ie\bar{u}(p^{'}) \Gamma^{\mu}(p,p^{'})u(p).
\label{eq:amplitude}
\end{equation}  
Here, this $\Gamma^{\mu}(p,p^{'})$ is called as the vertex 
function which takes care of the possible interactions originating 
through radiative corrections which can be  written as a sum of
form factors \cite{Lee:1977tib} like,
\begin{equation}
\Gamma^{\mu}(q) = \gamma^{\mu}F_1(q^2) + \frac{i\sigma^{\mu\nu}q_{\nu}}
{2m_{\mu}}F_2(q^2) + \frac{i\gamma^5 \sigma^{\mu\nu}q_{\nu}}{2m_{\mu}}F_3(q^2).
\label{eq:vertex}
\end{equation}
	
Here, $F_1, F_2$, and $F_3$ are the form factors that give the electric
charge, magnetic moment, and electric dipole moment respectively. And $q$ 
is the photon momentum, $q=(p-p')$. Here, the Majorana nature of fermions 
can effect the vertex function with the presence of charge conjugation
operator. As for the Majorana fermions particle and antiparticle states
are the same, it is expressed as, $\psi^c = C\bar{\psi^T}$. So, following 
the amplitude given in Eq. (\ref{eq:amplitude}), the coefficients of
form factors take the form after applying the charge conjugation 
operator, like
\begin{equation}
C\gamma_{\mu}^TC^{-1} = -\gamma_{\mu}, \quad C\sigma_{\mu\nu}C^{-1} 
= -\sigma_{\mu\nu}, \quad C\sigma_{\mu\nu}\gamma_5C^{-1} = 
-\sigma_{\mu\nu}\gamma_5.  
\end{equation}    
	
Under this transformations the vertex function in
Eq.(\ref{eq:vertex}) becomes,
\begin{equation}
\Gamma^{\mu}(q) = -\gamma^{\mu}F_1(q^2) - \frac{i\sigma^{\mu\nu}
q_{\nu}}{2m_{\mu}}F_2(q^2) - \frac{i\gamma^5 \sigma^{\mu\nu}q_{\nu}}
{2m_{\mu}}F_3(q^2).
\label{eq:mvertex}
\end{equation}
	
Now, extracting the form factor $F_2(q^2)$, the anomalous 
magnetic moment for muon term can be derived for the 
$(q^2 = 0)$ value,
\begin{equation}
a_{\mu} = \frac{g-2}{2} = F_{2}(q^2 = 0).
\end{equation}
\subsection{Feynman parametrization}
\label{app:b}
The Feynman parametrization is a technique developed by 
Richard Feynman. Where the multiplication of two numbers is 
expressed in the form of a finite integral. The simplest 
example is, 
\begin{equation}
\frac{1}{ab} = \int_{0}^{1}dx\int_{0}^{1}\frac{\delta(1-x-y)dy}
{[ax + by]^2}.
\end{equation} 
For, three numbers the integral stands as,
\begin{equation}
\begin{split}
\frac{1}{abc} &= 2\int_{0}^{1}dx\int_{0}^{1}dy\int_{0}^{1}
\frac{\delta(1-x-y-z)dz}{[ax + by + cz]^3},\\& =2\int_{0}^{1}dx\int_{0}^{1-x}
\frac{dy}{[ax + by + c(1-x-y)]^3}.
\end{split}
\end{equation}
In Eq.(\ref{eq:vf2}), this relation is used to simplify the 
denominator term, where, $a = (k'^2 - M_i^2 + i\epsilon)$,
$b = (k^{2} - M_j^2 + i\epsilon)$ and $c = ((p-k)^2 - m_{W_1}^2 
+ i\epsilon)$. Using, the expressions for a,b and c, the integral 
for the denominator term is expressed as,
\begin{equation}
\begin{split}
&\hspace{3cm}\frac{1}{(k'^2 - M_i^2 + i\epsilon)(k^{2} - 
M_j^2 + i\epsilon)((p-k)^2 - m_{W_1}^2 + i\epsilon)} \\&= 
2\int_{0}^{1}dx\int_{0}^{1-x}
\frac{dy}{[(k'^2 - M_i^2 + i\epsilon)x + (k^{2} - M_j^2 + 
i\epsilon)y + ((p-k)^2 - m_{W_1}^2 + i\epsilon)(1-x-y)]^3}.
 \end{split}
\end{equation}
And for the numerator part, the simplification is done 
using the Clifford algebra.
	
This  idea can be generalized for n'th variable which is
given by,
\begin{equation}
\frac{1}{a_1a_2...a_n} = (n-1)!\int_{0}^{1}dx_1\int_{0}^{1}dx_2...
\int_{0}^{1}dx_n \frac{\delta(1-x_1 - x_2 - ... -x_n)}{[a_1x_1 + 
a_2x_2 + ... + a_nx_n]^n}.
\end{equation}
This parametrization is extremely helpful for calculations that 
include multiple propagators. 

\bibliographystyle{plain}

\begin{thebibliography}{10}

\bibitem{Mohapatra:1974gc}
R.~N. Mohapatra and Jogesh~C. Pati.
\newblock {A Natural Left-Right Symmetry}.
\newblock {\em Phys. Rev. D}, 11:2558, 1975.

\bibitem{Senjanovic:1975rk}
G.~Senjanovic and Rabindra~N. Mohapatra.
\newblock {Exact Left-Right Symmetry and Spontaneous Violation of Parity}.
\newblock {\em Phys. Rev. D}, 12:1502, 1975.

\bibitem{Minkowski:1977sc}
Peter Minkowski.
\newblock {$\mu \to e\gamma$ at a Rate of One Out of $10^{9}$ Muon Decays?}
\newblock {\em Phys. Lett. B}, 67:421--428, 1977.

\bibitem{Ramond:1979py}
Pierre Ramond.
\newblock {The Family Group in Grand Unified Theories}.
\newblock In {\em {International Symposium on Fundamentals of Quantum Theory
  and Quantum Field Theory}}, 2 1979.

\bibitem{Gell-Mann:1979vob}
Murray Gell-Mann, Pierre Ramond, and Richard Slansky.
\newblock {Complex Spinors and Unified Theories}.
\newblock {\em Conf. Proc. C}, 790927:315--321, 1979.

\bibitem{Sawada:1979dis}
Osamu Sawada and Akio Sugamoto, editors.
\newblock {\em {Proceedings: Workshop on the Unified Theories and the Baryon
  Number in the Universe}: {Tsukuba, Japan, February 13-14, 1979}}, Tsukuba,
  Japan, 1979. Natl.Lab.High Energy Phys.

\bibitem{Levy:1980ws}
{\em {QUARKS AND LEPTONS. PROCEEDINGS, SUMMER INSTITUTE, CARGESE, FRANCE, JULY
  9-29, 1979}}, volume~61, 1980.

\bibitem{Magg:1980ut}
M.~Magg and C.~Wetterich.
\newblock {Neutrino Mass Problem and Gauge Hierarchy}.
\newblock {\em Phys. Lett. B}, 94:61--64, 1980.

\bibitem{Lazarides:1980nt}
George Lazarides, Q.~Shafi, and C.~Wetterich.
\newblock {Proton Lifetime and Fermion Masses in an SO(10) Model}.
\newblock {\em Nucl. Phys. B}, 181:287--300, 1981.

\bibitem{Mohapatra:1980yp}
Rabindra~N. Mohapatra and Goran Senjanovic.
\newblock {Neutrino Masses and Mixings in Gauge Models with Spontaneous Parity
  Violation}.
\newblock {\em Phys. Rev. D}, 23:165, 1981.

\bibitem{Schechter:1980gr}
J.~Schechter and J.~W.~F. Valle.
\newblock {Neutrino Masses in SU(2) x U(1) Theories}.
\newblock {\em Phys. Rev. D}, 22:2227, 1980.

\bibitem{Zupan:2019uoi}
Jure Zupan.
\newblock {Introduction to flavour physics}.
\newblock {\em CERN Yellow Rep. School Proc.}, 6:181--212, 2019.

\bibitem{Nemevsek:2012iq}
Miha Nemevsek, Goran Senjanovic, and Vladimir Tello.
\newblock {Connecting Dirac and Majorana Neutrino Mass Matrices in the Minimal
  Left-Right Symmetric Model}.
\newblock {\em Phys. Rev. Lett.}, 110(15):151802, 2013.

\bibitem{Senjanovic:2018xtu}
Goran Senjanovic and Vladimir Tello.
\newblock {Disentangling the seesaw mechanism in the minimal left-right
  symmetric model}.
\newblock {\em Phys. Rev. D}, 100(11):115031, 2019.

\bibitem{Casas:2001sr}
J.~A. Casas and A.~Ibarra.
\newblock {Oscillating neutrinos and $\mu \to e, \gamma$}.
\newblock {\em Nucl. Phys. B}, 618:171--204, 2001.

\bibitem{Akhmedov:2006de}
Evgeny~K. Akhmedov and M.~Frigerio.
\newblock {Interplay of type I and type II seesaw contributions to neutrino
  mass}.
\newblock {\em JHEP}, 01:043, 2007.

\bibitem{Hosteins:2006ja}
Pierre Hosteins, Stephane Lavignac, and Carlos~A. Savoy.
\newblock {Quark-Lepton Unification and Eight-Fold Ambiguity in the Left-Right
  Symmetric Seesaw Mechanism}.
\newblock {\em Nucl. Phys. B}, 755:137--163, 2006.

\bibitem{Banerjee:2023aro}
Vivek Banerjee and Sasmita Mishra.
\newblock {Interplay of type-I and type-II seesaw in neutrinoless double beta
  decay in left-right symmetric model}.
\newblock {\em Nucl. Phys. B}, 1006:116623, 2024.

\bibitem{Bernstein:1963qh}
Jeremy Bernstein, Malvin Ruderman, and Gerald Feinberg.
\newblock {Electromagnetic Properties of the neutrino}.
\newblock {\em Phys. Rev.}, 132:1227--1233, 1963.

\bibitem{Pal:1981rm}
Palash~B. Pal and Lincoln Wolfenstein.
\newblock {Radiative Decays of Massive Neutrinos}.
\newblock {\em Phys. Rev. D}, 25:766, 1982.

\bibitem{PhysRevD.34.3444}
Utpal Chattopadhyay and Palash~B. Pal.
\newblock Radiative neutrino decay in left-right models.
\newblock {\em Phys. Rev. D}, 34:3444--3448, Dec 1986.

\bibitem{Babu:1991zh}
K.~S. Babu.
\newblock {Neutrino masses and magnetic moments}.
\newblock In {\em {4th Mexican School of Particles and Fields}}, 6 1991.

\bibitem{Akhmedov:1993ta}
Evgeny~K. Akhmedov, S.~T. Petcov, and A.~Yu. Smirnov.
\newblock {Pontecorvo's original oscillations revisited}.
\newblock {\em Phys. Lett. B}, 309:95--102, 1993.

\bibitem{Mikheev:1986wj}
S.~P. Mikheev and A.~Yu. Smirnov.
\newblock {Resonant amplification of neutrino oscillations in matter and solar
  neutrino spectroscopy}.
\newblock {\em Nuovo Cim. C}, 9:17--26, 1986.

\bibitem{Smirnov:1991ia}
A.~Yu. Smirnov.
\newblock {The Geometrical phase in neutrino spin precession and the solar
  neutrino problem}.
\newblock {\em Phys. Lett. B}, 260:161--164, 1991.

\bibitem{Wolfenstein:1977ue}
L.~Wolfenstein.
\newblock {Neutrino Oscillations in Matter}.
\newblock {\em Phys. Rev. D}, 17:2369--2374, 1978.

\bibitem{Giunti:2014ixa}
Carlo Giunti and Alexander Studenikin.
\newblock {Neutrino electromagnetic interactions: a window to new physics}.
\newblock {\em Rev. Mod. Phys.}, 87:531, 2015.

\bibitem{Fujikawa:1980yx}
Kazuo Fujikawa and Robert Shrock.
\newblock {The Magnetic Moment of a Massive Neutrino and Neutrino Spin
  Rotation}.
\newblock {\em Phys. Rev. Lett.}, 45:963, 1980.

\bibitem{Babu:1989wn}
K.~S. Babu and R.~N. Mohapatra.
\newblock {Model for Large Transition Magnetic Moment of the $\nu_e$}.
\newblock {\em Phys. Rev. Lett.}, 63:228, 1989.




\bibitem{Borexino:2017fbd}
M.~Agostini et~al.
\newblock {Limiting neutrino magnetic moments with Borexino Phase-II solar
  neutrino data}.
\newblock {\em Phys. Rev. D}, 96(9):091103, 2017.



\bibitem{TEXONO:2006xds}
H.~T. Wong et~al.
\newblock {A Search of Neutrino Magnetic Moments with a High-Purity Germanium
  Detector at the Kuo-Sheng Nuclear Power Station}.
\newblock {\em Phys. Rev. D}, 75:012001, 2007.





\bibitem{Liu:1989me}
Jiang Liu.
\newblock {Neutrino Magnetic Moment in the SU(3)-l X U(1) Model}.
\newblock {\em Phys. Lett. B}, 225:148--152, 1989.

\bibitem{Tarazona:2015drz}
Carlos~G. Tarazona, Rodolfo~A. Diaz, John Morales, and Andr\'es Castillo.
\newblock {Phenomenology of the new physics coming from 2HDMs to the neutrino
  magnetic dipole moment}.
\newblock {\em Int. J. Mod. Phys. A}, 32(10):1750050, 2017.

\bibitem{Boyarkin:2008zz}
O.~M. Boyarkin, G.~G. Boyarkina, and V.~V. Makhnach.
\newblock {$(g-2)_{\mu}$ anomaly within the left-right symmetric model}.
\newblock {\em Phys. Rev. D}, 77:033004, 2008.

\bibitem{Aparici:2009fh}
Alberto Aparici, Kyungwook Kim, Arcadi Santamaria, and Jose Wudka.
\newblock {Right-handed neutrino magnetic moments}.
\newblock {\em Phys. Rev. D}, 80:013010, 2009.

\bibitem{Aparici:2009oua}
Alberto Aparici, Arcadi Santamaria, and Jose Wudka.
\newblock {A model for right-handed neutrino magnetic moments}.
\newblock {\em J. Phys. G}, 37:075012, 2010.

\bibitem{Kikuchi:2008ki}
Tatsuru Kikuchi.
\newblock {Enhanced production of TeV right-handed neutrinos through the large
  magnetic moment}.
\newblock {\em Phys. Lett. B}, 671:272--274, 2009.

\bibitem{Barducci:2022gdv}
Daniele Barducci, Enrico Bertuzzo, Marco Taoso, and Claudio Toni.
\newblock {Probing right-handed neutrinos dipole operators}.
\newblock {\em JHEP}, 03:239, 2023.

\bibitem{Chun:2024mus}
Eung~Jin Chun, Sanjoy Mandal, and Rojalin Padhan.
\newblock {Collider imprints of a right handed neutrino magnetic moment
  operator}.
\newblock {\em Phys. Rev. D}, 109(11):115002, 2024.

\bibitem{Deshpande:1990ip}
N.~G. Deshpande, J.~F. Gunion, Boris Kayser, and Fredrick~I. Olness.
\newblock {Left-right symmetric electroweak models with triplet Higgs}.
\newblock {\em Phys. Rev. D}, 44:837--858, 1991.

\bibitem{Senjanovic:1978ev}
Goran Senjanovic.
\newblock {Spontaneous Breakdown of Parity in a Class of Gauge Theories}.
\newblock {\em Nucl. Phys. B}, 153:334--364, 1979.

\bibitem{Mohapatra:1979ia}
Rabindra~N. Mohapatra and Goran Senjanovic.
\newblock {Neutrino Mass and Spontaneous Parity Nonconservation}.
\newblock {\em Phys. Rev. Lett.}, 44:912, 1980.

\bibitem{Akhmedov:2006yp}
Evgeny~K. Akhmedov, Mattias Blennow, Tomas Hallgren, Thomas Konstandin, and  Tommy Ohlsson.
\newblock {Stability and leptogenesis in the left-right symmetric seesaw
  mechanism}.
\newblock {\em JHEP}, 04:022, 2007.

\bibitem{Barry:2013xxa}
James Barry and Werner Rodejohann.
\newblock {Lepton number and flavour violation in TeV-scale left-right
  symmetric theories with large left-right mixing}.
\newblock {\em JHEP}, 09:153, 2013.

\bibitem{Rothstein:1990qx}
I.~Z. Rothstein.
\newblock {Renormalization group analysis of the minimal left-right symmetric
  model}.
\newblock {\em Nucl. Phys. B}, 358:181--194, 1991.

\bibitem{Staub:2013tta}
Florian Staub.
\newblock {SARAH 4 : A tool for (not only SUSY) model builders}.
\newblock {\em Comput. Phys. Commun.}, 185:1773--1790, 2014.

\bibitem{Giudice:2003jh}
G.~F. Giudice, A.~Notari, M.~Raidal, A.~Riotto, and A.~Strumia.
\newblock {Towards a complete theory of thermal leptogenesis in the SM and
  MSSM}.
\newblock {\em Nucl. Phys. B}, 685:89--149, 2004.

\bibitem{Antusch:2003kp}
Stefan Antusch, J\"orn Kersten, Manfred Lindner, and Michael Ratz.
\newblock {Running neutrino masses, mixings and CP phases: Analytical results
  and phenomenological consequences}.
\newblock {\em Nucl. Phys. B}, 674:401--433, 2003.

\bibitem{Shrock:1982sc}
Robert~E. Shrock.
\newblock {Electromagnetic Properties and Decays of Dirac and Majorana
  Neutrinos in a General Class of Gauge Theories}.
\newblock {\em Nucl. Phys. B}, 206:359--379, 1982.

\bibitem{Czakon:2002vn}
M.~Czakon, J.~Gluza, and J.~Hejczyk.
\newblock {The Minimal left-right symmetric model and radiative corrections to
  the muon decay}.
\newblock {\em Nucl. Phys. B Proc. Suppl.}, 116:230--234, 2003.

\bibitem{Jegerlehner:2009ry}
Fred Jegerlehner and Andreas Nyffeler.
\newblock {The Muon g-2}.
\newblock {\em Phys. Rept.}, 477:1--110, 2009.

\bibitem{Aoyama:2020ynm}
T.~Aoyama et~al.
\newblock {The anomalous magnetic moment of the muon in the Standard Model}.
\newblock {\em Phys. Rept.}, 887:1--166, 2020.

\bibitem{Muong-2:2021ojo}
B.~Abi et~al.
\newblock {Measurement of the Positive Muon Anomalous Magnetic Moment to 0.46
  ppm}.
\newblock {\em Phys. Rev. Lett.}, 126(14):141801, 2021.

\bibitem{Blum:2019ugy}
Thomas Blum, Norman Christ, Masashi Hayakawa, Taku Izubuchi, Luchang Jin,
  Chulwoo Jung, and Christoph Lehner.
\newblock {Hadronic Light-by-Light Scattering Contribution to the Muon
  Anomalous Magnetic Moment from Lattice QCD}.
\newblock {\em Phys. Rev. Lett.}, 124(13):132002, 2020.

\bibitem{DellaMorte:2017dyu}
M.~Della~Morte, A.~Francis, V.~G\"ulpers, G.~Herdo\'\i{}za, G.~von Hippel,
  H.~Horch, B.~J\"ager, H.~B. Meyer, A.~Nyffeler, and H.~Wittig.
\newblock {The hadronic vacuum polarization contribution to the muon $g-2$ from
  lattice QCD}.
\newblock {\em JHEP}, 10:020, 2017.

\bibitem{FermilabLatticeHPQCD:2023jof}
Alexei Bazavov et~al.
\newblock {Light-quark connected intermediate-window contributions to the muon
  g-2 hadronic vacuum polarization from lattice QCD}.
\newblock {\em Phys. Rev. D}, 107(11):114514, 2023.

\bibitem{DiLuzio:2021uty}
Luca Di~Luzio, Antonio Masiero, Paride Paradisi, and Massimo Passera.
\newblock {New physics behind the new muon g-2 puzzle?}
\newblock {\em Phys. Lett. B}, 829:137037, 2022.

\bibitem{Queiroz:2014zfa}
Farinaldo~S. Queiroz and William Shepherd.
\newblock {New Physics Contributions to the Muon Anomalous Magnetic Moment: A
  Numerical Code}.
\newblock {\em Phys. Rev. D}, 89(9):095024, 2014.

\bibitem{PhysRevD.16.2256}
T.~Goldman and G.~J. Stephenson.
\newblock Limits on the mass of the muon neutrino in the absence of
  muon-lepton-number conservation.
\newblock {\em Phys. Rev. D}, 16:2256--2259, Oct 1977.

\bibitem{Petcov:1976ff}
S.~T. Petcov.
\newblock {The Processes $\mu \rightarrow e + \gamma, \mu \rightarrow e +
  \overline{e}, \nu' \rightarrow \nu + \gamma$ in the Weinberg-Salam Model with
  Neutrino Mixing}.
\newblock {\em Sov. J. Nucl. Phys.}, 25:340, 1977.
\newblock [Erratum: Sov.J.Nucl.Phys. 25, 698 (1977), Erratum: Yad.Fiz. 25, 1336
  (1977)].

\bibitem{PhysRev.76.769}
R.~P. Feynman.
\newblock Space-time approach to quantum electrodynamics.
\newblock {\em Phys. Rev.}, 76:769--789, Sep 1949.

\bibitem{Esteban:2024eli}
Ivan Esteban, M.~C. Gonzalez-Garcia, Michele Maltoni, Ivan Martinez-Soler,
  Jo\~ao~Paulo Pinheiro, and Thomas Schwetz.
\newblock {NuFit-6.0: updated global analysis of three-flavor neutrino
  oscillations}.
\newblock {\em JHEP}, 12:216, 2024.

\bibitem{ParticleDataGroup:2020ssz}
P.~A. Zyla et~al.
\newblock {Review of Particle Physics}.
\newblock {\em PTEP}, 2020(8):083C01, 2020.

\bibitem{ATLAS:2012ak}
Georges Aad et~al.
\newblock {Search for heavy neutrinos and right-handed $W$ bosons in events
  with two leptons and jets in $pp$ collisions at $\sqrt{s}=7$ TeV with the
  ATLAS detector}.
\newblock {\em Eur. Phys. J. C}, 72:2056, 2012.

\bibitem{CMS:2012zv}
Serguei Chatrchyan et~al.
\newblock {Search for Heavy Neutrinos and $W_R$ Bosons with Right-Handed
  Couplings in a Left-Right Symmetric Model in pp Collisions at $\sqrt{s}$ = 7
  TeV}.
\newblock {\em Phys. Rev. Lett.}, 109:261802, 2012.

\bibitem{Gozdz:2014pla}
Marek G\'o\'zd\'z and Wies\l{}aw~A. Kami\'nski.
\newblock {$0\nu2\beta$ decay and neutrino magnetic moment}.
\newblock {\em Phys. Rev. D}, 89(11):113005, 2014.

\bibitem{Gozdz:2014gna}
Marek G\'o\'zd\'z and Wies\l{}aw~A. Kami\'nski.
\newblock {Neutrinoless double beta decay mediated by the neutrino magnetic
  moment}.
\newblock {\em Acta Phys. Polon. B}, 47:1245, 2016.

\bibitem{Kotila:2012zza}
J.~Kotila and F.~Iachello.
\newblock {Phase space factors for double-$\beta$ decay}.
\newblock {\em Phys. Rev. C}, 85:034316, 2012.

\bibitem{Dueck:2011hu}
Alexander Dueck, Werner Rodejohann, and Kai Zuber.
\newblock {Neutrinoless Double Beta Decay, the Inverted Hierarchy and Precision
  Determination of theta(12)}.
\newblock {\em Phys. Rev. D}, 83:113010, 2011.

\bibitem{Faessler:2011rv}
Amand Faessler, G.~L. Fogli, E.~Lisi, A.~M. Rotunno, and F.~Simkovic.
\newblock {Multi-Isotope Degeneracy of Neutrinoless Double Beta Decay
  Mechanisms in the Quasi-Particle Random Phase Approximation}.
\newblock {\em Phys. Rev. D}, 83:113015, 2011.

\bibitem{Faessler:2011qw}
Amand Faessler, A.~Meroni, S.~T. Petcov, F.~Simkovic, and J.~Vergados.
\newblock {Uncovering Multiple $CP$-Nonconserving Mechanisms of $(\beta
  \beta)_{0\nu}$ Decay}.
\newblock {\em Phys. Rev. D}, 83:113003, 2011.

\bibitem{KamLAND-Zen:2022tow}
S.~Abe et~al.
\newblock {Search for the Majorana Nature of Neutrinos in the Inverted Mass
  Ordering Region with KamLAND-Zen}.
\newblock {\em Phys. Rev. Lett.}, 130(5):051801, 2023.

\bibitem{Shtembari:2022pfs}
Lolian Shtembari.
\newblock {Final results from GERDA: a neutrinoless double beta decay search}.
\newblock {\em PoS}, ICHEP2022:614, 2022.

\bibitem{nEXO:2017nam}
J.~B. Albert et~al.
\newblock {Sensitivity and Discovery Potential of nEXO to Neutrinoless Double
  Beta Decay}.
\newblock {\em Phys. Rev. C}, 97(6):065503, 2018.

\bibitem{LEGEND:2017cdu}
N.~Abgrall et~al.
\newblock {The Large Enriched Germanium Experiment for Neutrinoless Double Beta
  Decay (LEGEND)}.
\newblock {\em AIP Conf. Proc.}, 1894(1):020027, 2017.

\bibitem{Armengaud:2019loe}
E.~Armengaud et~al.
\newblock {The CUPID-Mo experiment for neutrinoless double-beta decay:
  performance and prospects}.
\newblock {\em Eur. Phys. J. C}, 80(1):44, 2020.

\bibitem{Dmitriev_2016}
Arkadiy~Nikholaevich Dmitriev.
\newblock A new look on the nature of the earth's magnetic field.
\newblock {\em IOP Conference Series: Earth and Environmental Science},
  44(2):022001, oct 2016.

\bibitem{Peskin:1995ev}
Michael~E. Peskin and Daniel~V. Schroeder.
\newblock {\em {An Introduction to quantum field theory}}.
\newblock Addison-Wesley, Reading, USA, 1995.

\bibitem{Czarnecki:2002nt}
Andrzej Czarnecki, William~J. Marciano, and Arkady Vainshtein.
\newblock {Refinements in electroweak contributions to the muon anomalous
  magnetic moment}.
\newblock {\em Phys. Rev. D}, 67:073006, 2003.
\newblock [Erratum: Phys.Rev.D 73, 119901 (2006)].

\bibitem{Bars:1972pe}
I.~Bars and M.~Yoshimura.
\newblock {Muon magnetic moment in a finite theory of weak and electromagnetic
  interaction}.
\newblock {\em Phys. Rev. D}, 6:374--376, 1972.

\bibitem{Lee:1977tib}
Benjamin~W. Lee and Robert~E. Shrock.
\newblock {Natural Suppression of Symmetry Violation in Gauge Theories: Muon -
  Lepton and Electron Lepton Number Nonconservation}.
\newblock {\em Phys. Rev. D}, 16:1444, 1977.
\end{thebibliography}

\end{document}